\documentclass{aa}

\usepackage{graphicx}
\usepackage{txfonts}
\usepackage{natbib}
\usepackage{float}
\usepackage[colorlinks=true, linkcolor=blue, citecolor=blue, urlcolor=blue]{hyperref}

\begin{document}

\title{Survival of molecular gas in cavities of transition disks}

\subtitle{I. CO}

\author{Simon Bruderer\inst{\ref{inst_mpe}}}

\institute{
Max-Planck-Institut f\"{u}r Extraterrestrische Physik, Giessenbachstrasse 1, 85748 Garching, Germany\label{inst_mpe}
}

\date{Accepted by A\&A, July 26th 2013}

\titlerunning{Survival of molecules in cavities of transition disks}
\authorrunning{S. Bruderer}

\offprints{Simon Bruderer,\\ \email{simonbruderer@gmail.com}}

\abstract
{Planet formation is closely related to the structure and dispersal of protoplanetary disks. A certain class of disks, called transition disks, exhibit cavities in dust images at scales of up to a few 10s of AU. The formation mechanism of the cavities is still unclear. The gas content of such cavities can be spatially resolved for the first time using the Atacama Large Millimeter/submillimeter Array (ALMA).} 
{To develop a new series of models to simulate the physical conditions and chemical abundances of the gas in cavities to address the question whether the gas is primarily atomic or molecular inside the dust free cavities exposed to intense UV radiation. Molecular/atomic line emission by carbon monoxide (CO), its isotopologues ($^{13}$CO, C$^{18}$O, C$^{17}$O, and $^{13}$C$^{18}$O) and related species ([\ion{C}{I}], [\ion{C}{II}], and [\ion{O}{I}]) is predicted for comparison with ALMA and the Herschel Space Observatory.} 
{We use a thermo-chemical model, which calculates the radiative transfer both in lines and the continuum, and solves for the chemical abundances and gas temperature. The model is based on our previous work, but includes several improvements. We study the dependence of CO abundances and lines on several parameters such as gas mass in the cavity, disk mass and luminosity of the star.} 
{The gas can remain in molecular form down to very low amounts of gas in the cavity ($\sim$1 \% of M$_{\rm Earth}$). Shielding of the stellar radiation by a dusty inner disk (``pre-transition disk'') allows CO to survive down to lower gas masses in the cavity. The column densities of H$_2$ and CO in the cavity scale almost linearly with the amount of gas in the cavity down to the mass where photodissociation becomes important. The main parameter for the CO emission from cavity is the gas mass. Other parameters such as the outer disk mass, bolometric luminosity, shape of the stellar spectrum or PAH abundance are less important. Since the CO pure rotational lines readily become optically thick, the CO isotopologues need to be observed in order to quantitatively determine the amount of gas in the cavity. Determining gas masses in the cavity from atomic lines ([\ion{C}{I}], [\ion{C}{II}], and [\ion{O}{I}]) is challenging.} 
{A wide range of gas masses in the cavity of transition disks ($\sim 4$ orders of magnitude) can be probed using combined observations of CO isotopologue lines with ALMA. Measuring the gas mass in the cavity will ultimately help to distinguish between different cavity formation theories.} 
\maketitle

%
%

\section{Introduction} \label{sec:intro}

Planets form in accretion disks around young stars. A detailed understanding of the evolution and dissipation of protoplanetary disks is thus crucial to constrain theories of planet formation (see \citealt{Armitage11}, for a recent review). An important stage of the evolution is the phase of gas dispersal which sets an end to the formation of (gas giant) planets. While early gas-rich protoplanetary disks around classical T Tauri stars and late, gas-poor debris disks have been studied in detail (\citealt{Wyatt08,Williams11} for recent reviews), the transitional phase between these two stages remains enigmatic.

The so-called \emph{transition disks} with ongoing evolution in their dust and/or gas structure have so far mostly been studied by their continuum emission. Their spectral energy distributions (SEDs) (e.g., \citealt{Strom89,Skrutskie90,Calvet02,Calvet05,dAlessio05,Espaillat07b,Espaillat08a,Merin10,Currie11,Cieza12,Romero12}) have strong excess over the stellar photosphere at $>$20 $\mu$m, but no or little excess at shorter wavelengths. This was interpreted with an absence of small and warm dust from the inner part of the disk, with hole sizes from $\sim5$ to $\sim50$ AU. Direct mapping with submillimeter interferometry (\citealt{Dutrey08,Brown08,Brown09}, \citealt{Andrews11}, hereafter A11, and \citealt{Mathews12}) directly reveal the larger of these dust holes.  A subclass of transition disks\footnote{This type of transition disks is sometimes called ``pre-transition'' disk. In our terminology, both are called transition disks, either with or without a dusty inner disk.} shows evidence of warm optically thick dust near the star in their SED (e.g., \citealt{Brown07,Espaillat10}), indicative of a dust gap rather than a dust hole.\\

In contrast, the gas content of transition disks has received much less attention and has been focused largely on H$\alpha$ studies probing the ionized gas accreting onto the star (e.g., \citealt{Cieza10,Cieza12,Romero12,Espaillat12}). Little is known about the neutral gas inside dust cavities and it is not even clear how to trace it. Can molecules actually survive inside dust cavities? In the absence of small dust grains, the molecules are not protected from the intense stellar UV radiation by dust shielding. Moreover, the formation of the most basic molecule, H$_2$, is limited in the absence of small dust grains. Nevertheless, pioneering observations at near-IR wavelengths have detected rovibrational emission of CO at 4.7 $\mu$m in some sources indicating the presence of hot molecular gas inside their dust cavities \citep{Salyk07, Salyk09, Salyk11b, Pontoppidan08a,Brittain07,Brown12}. In a few cases, spectrally and spatially resolved lines allow kinematic modeling and point to different inner radii of the CO emission and the dust cavity (\citealt{Pontoppidan08a}; \citealt{Brown12}). However, these lines only trace hot gas ($\gtrsim 300$ K) in which the vibrational levels are excited and they cannot be used to constrain the total amount of gas, thought to be at lower temperatures ($\lesssim$ few 100 K). Millimeter/submillimeter observations of pure rotational lines of CO are much better suited to constrain the gas mass in the cavity, but have so far been plagued by (too) low angular resolution and sensitivity of current (sub)millimeter interferometers (e.g. \citealt{Dutrey08}, \citealt{Lyo11}).

The Atacama Large Millimeter/submillimeter Array (ALMA\footnote{www.almaobservatory.org}) allows the cold gas in cavities of transition disks to be studied for the first time (\citealt{Casassus13}; \citealt{vdMarel13}). Low-$J$ CO lines observed by ALMA at high angular resolution have several advantages over the near-IR lines. The millimeter lines have low critical densities and are thus in local thermal equilibrium (LTE) which makes the excitation calculation simple. The level energies are low so the bulk of the mass, where near-IR lines are not excited, can be probed. Due to their long wavelengths, the lines are less prone to dust absorption than near/mid-IR lines as e.g. the rovibrational lines of CO at 4.7 $\mu$m or the optical [OI] 6300 \AA~ line (e.g., \citealt{Acke06}). Isotopologues of CO can be observed, which allows a wide range of column densities to be measured if the lines of the main isotopologue are optically thick. If gas is present in the cavity, but the conditions do not allow the formation of CO, carbon is likely in the form of neutral (C) or ionized atomic carbon (C$^+$) and oxygen as neutral O. Neutral carbon has lines at submillimeter wavelengths ([\ion{C}{I}] $^3$P$_{2}$-$^3$P$_{1}$ at 370 $\mu$m/809 GHz, and [\ion{C}{I}] $^3$P$_{1}$-$^3$P$_{0}$ at 609 $\mu$m/492 GHz) and can thus also be observed by ALMA. Ionized carbon has the fine structure line [\ion{C}{II}] $^2$P$_{3/2}$-$^2$P$_{1/2}$ at 158 $\mu$m and neutral oxygen [\ion{O}{I}] $^3$P$_{1}$-$^3$P$_{2}$ at 63 $\mu$m and [\ion{O}{I}] $^3$P$_{0}$-$^3$P$_{1}$ at 145 $\mu$m which can be observed by the {\it Herschel Space Observatory}.

Even if the millimeter lines of CO are detected, however, the observed emission cannot be directly translated into a gas mass since it is unclear a priori to what extent the gas is largely molecular or atomic, i.e., whether the molecules can actually survive the strong UV radiation  inside the dust free cavity. To determine the amount of gas that is present inside a dust cavity from observations, the physical and chemical structure of the gas needs to be modeled. We present here a set of thermo-chemical models to explore under which conditions molecules like CO are indeed present inside dust cavities and can be used to trace the gas mass. This, in turn, is relevant to constrain different mechanisms that have been proposed for the dust cavity formation.

Modeling gas line emission from protoplanetary disks is a complex task, because the line emission depends on various parameters (e.g. abundance, gas temperature, collision partner density, velocity structure, \ldots). A comprehensive model should thus calculate these physical and chemical parameters self-consistently for the whole disk structure. Thermo-chemical models of protoplanetary disks solve the problem in a similar way as classical PDR models (e.g. \citealt{Tielens85}): Using the local UV radiation field, a chemical network simulation calculates the abundance of various atoms and molecules. The abundances feed into the calculation of heating and cooling rates to obtain the gas temperature. Since the chemical network simulation depends on the gas temperature, the problem needs to be solved iteratively. Various thermo-chemical models of protoplanetary disks have been presented in the past (e.g. \citealt{Kamp01,Glassgold04,Gorti04,Jonkheid04,Nomura05,Aikawa06,Woitke09,Woods09,Ercolano09a,Bruderer12}). With the exception of \cite{Jonkheid06}, who modeled the molecular emission of one particular transition disk, most modeling effort focused on younger, gas rich disks rather than transition disks. \citet{Cleeves11} have also modeled a transition disk, but they focus on the outer disk and assume the cavity to be gas free.

In this work, we use our thermo-chemical model, which was presented together with benchmark tests in \cite{Bruderer12}, to calculate a grid of transition disk models. The goal of this work is to predict observables (e.g. line fluxes or images) of a transition disk with a given amount of dust/gas inside the cavity. We thus choose to use a simple approach which is based on a parameterized density structure. This structure was found by A11 to simultaneously fit the SED, submillimeter images and visibilities of twelve  transition disks. Besides being based on observational foundations, another advantage of this approach is that one can relate the amount gas in a generic way to predicted/observed line emission, without referring to one particular dust removal scenario. We put the initial focus on carbon monoxide (CO), atomic carbon (C, C$^+$) and oxygen (O) and will discuss other species later. In an extension to this work, we plan to couple our model to the dust evolution models by \cite{Birnstiel10,Birnstiel11,Birnstiel12} and \cite{Pinilla12}, in order to study the grain growth and planetary clearing scenario in more detail.

The paper is organized as follows: In Sect. \ref{sec:model} we summarize the model used in this work and explain the free parameters. The parameters chosen and caveats of the work are discussed next. The following sections present the model results: We first show the physical structure (e.g. density, temperature) of one representative model in Sect. \ref{sec:rep_physstructure}. The abundance structure of CO, H$_2$, C and C$^+$ depending on the key parameters are then discussed in Sect. \ref{sec:survival}. In Sect. \ref{sec:molemission} we discuss the CO pure rotational line emission and line opacity effects. The implications of our models to use rotational lines of CO and its isotopologues as tracers of the gas mass in the gap are discussed in Sect. \ref{sec:discussions}. The paper ends with a conclusion section.

%
%

\section{Model} \label{sec:model}

We use the radiative transfer, chemistry and thermal-balance model introduced by \cite{Bruderer12} (hereafter BR12). For the current application, we have implemented several improvements and minor changes, which are described in Appendix \ref{sec:app_mod}. Details of the model and results of benchmark tests are reported in BR12, however a brief outline is given here.

Starting from an input density structure, the model solves the continuum radiative transfer problem using a 3D Monte-Carlo method to obtain the dust temperature and local radiation field for UV to millimeter wavelengths. Non-isotropic scattering is accounted for. In a next step, a chemical network simulation is used to obtain the chemical composition. The reaction network is based on a subset of the UMIST 06 network (\citealt{Woodall07}) and consists of $\sim 110$ species and $\sim 1500$ reactions. Freeze-out, thermal and non-thermal evaporation and some basic grain-surface reactions of species hydrogenating on the grain surface (e.g. g:O $\rightarrow$ g:OH $\rightarrow$ g:H$_2$O, see \citealt{Visser11}) are included as well as formation of H$_2$ and CH$^+$ on the surface of PAHs (\citealt{Jonkheid06}). Photodissociation rates are obtained using the molecular cross-section (\citealt{vanDishoeck06c}\footnote{http://home.strw.leidenuniv.nl/\~{}ewine/photo/}) and the FUV intensity from the continuum transfer calculation, so that the effect of different central stars can be investigated. X-ray ionization and the effect of hot (vibrationally excited) H$_2^*$ are also accounted for. 

The chemical abundances are then used as input for a non-LTE excitation calculation of the main cooling atoms and molecules (e.g. O, C$^+$, CO, \ldots). An escape probability method is used and pumping by the dust continuum is accounted for. Molecular data for collisions and radiative rates are mainly taken from the LAMDA database (\citealt{Schoier05}). The gas temperature is obtained from the balance between heating processes (e.g. photoelectric heating on PAHs or small grains) and cooling processes (e.g. by line radiation or gas-grain collisions) similar to classical PDR models (e.g. \citealt{Tielens85,Sternberg89,Kaufman99,Meijerink05a}). Since the chemical composition and the molecular excitation depend on the gas temperature, the problem has to be solved iteratively. Once a solution is found, spectral image cubes are derived using a raytracer.

\subsection{Parameters of the model} \label{sec:physstruct}

\subsubsection{Density structure}

The density structure used for our models follows the simple parametric prescription proposed by A11. The structure is motivated by a simple viscous accretion disk model with the viscosity not changing with time and spatially distributed as $\nu \propto R^\gamma$ (\citealt{LyndenBell74}, \citealt{Hartmann98}). Dust and gas are significantly depleted in an inner cavity. A11 use the density structure to perform a simultaneous SED, submillimeter image and visibility fitting and find good agreement for all of the twelve  studied transition disks. They note that several parameters are degenerate and also depend on the assumed dust properties. Direct observational constraints on the radial or vertical structure of the gas are difficult to obtain (e.g. \citealt{Brown12}). We thus restrict our work to a study of the dependence on certain parameters. In the following, we refer to dust as a small $<$ cm size population of dust grains which contribute considerably to the opacity at UV to millimeter wavelengths.

\begin{figure*}[htb]
\sidecaption
\includegraphics[width=0.8\hsize]{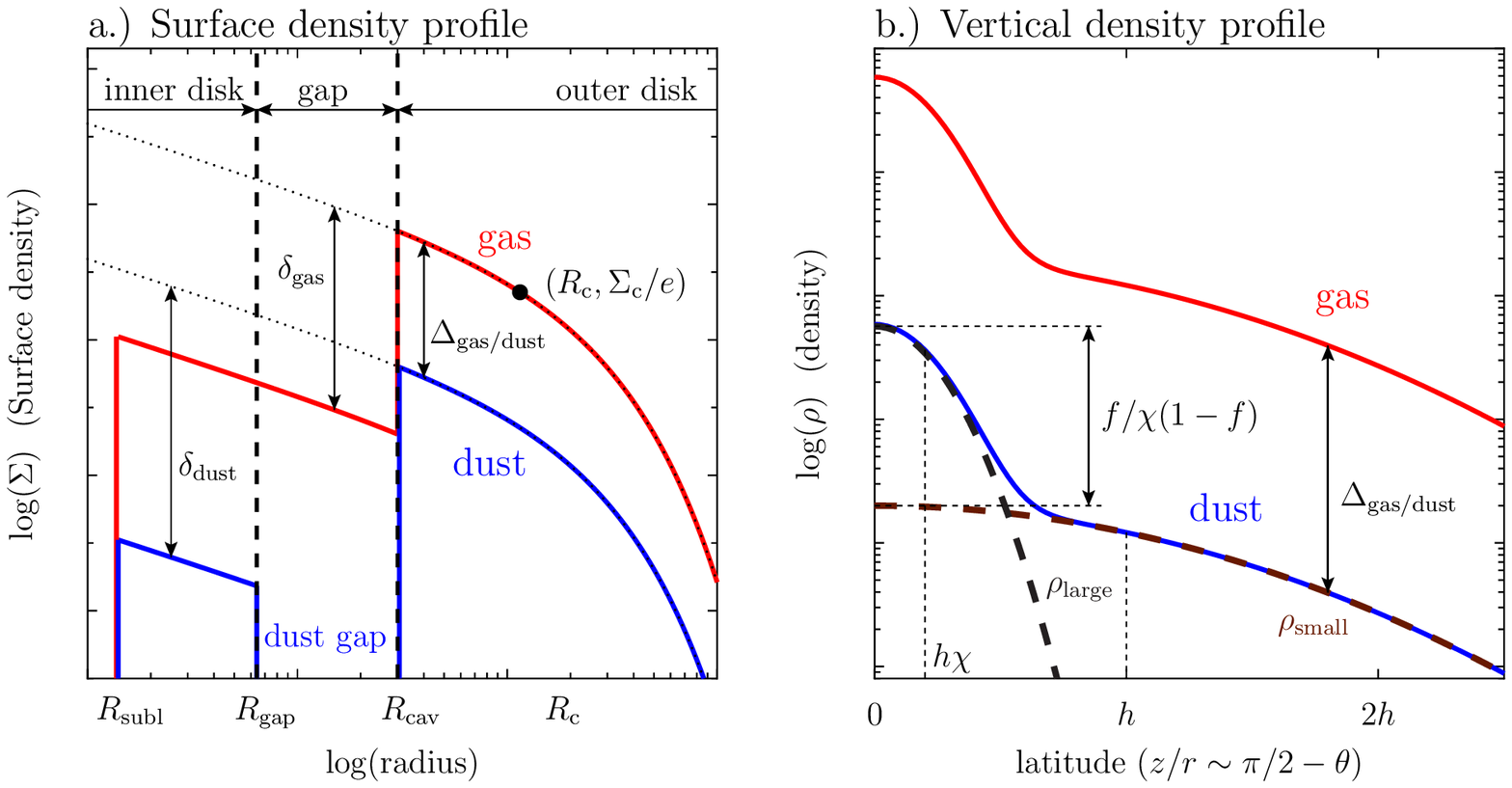}
\caption{Schematic view of the physical structure used in this work and definition of the main parameters. \textbf{a)} Surface density structure \textbf{b)} Vertical density structure.}
\label{fig:densstruct}
\end{figure*}

Figure \ref{fig:densstruct} shows the surface density and vertical structure used in our work. The surface density (Figure \ref{fig:densstruct}a) follows
\begin{equation}
\Sigma_{\rm gas} = \Sigma_c \left( \frac{R}{R_c} \right)^{-\gamma} \exp\left[- \left( \frac{R}{R_c} \right)^{2-\gamma} \right] \ .
\end{equation}
In the outer disk $R > R_{\rm cav}$ a gas-to-dust ratio $\Delta_{\rm gas/dust}=\Sigma_{\rm gas} / \Sigma_{\rm dust}$ is used. For $R \leqslant R_{\rm cav}$, the gas surface density is lowered by a factor of $\delta_{\rm gas}$ with respect to the outer disk and by $\delta_{\rm dust}$ for the dust. This remaining dust is required to reproduce the near-IR excess observed toward a class of transition disks (\citealt{Brown07,Espaillat10}; A11). Within the dust sublimation radius $R_{\rm subl} \sim 0.07 \sqrt{L_*/L_\odot}$ AU (assuming a dust sublimation temperature of $\sim$1500 K; \citealt{Dullemond01}), both dust and gas surface density are set to zero. Between the $R_{\rm gap} \geqslant R_{\rm subl}$ and $R_{\rm cav}$, dust is assumed to be absent. In this paper, we refer to the cavity as the region inside $R_{\rm cav}$ and the gap as the region between $R_{\rm gap}$ and $R_{\rm cav}$.

The vertical distribution of densities (Figure \ref{fig:densstruct}b) follows a Gaussian with scale height angle $h=h_c \left( R / R_c  \right)^\psi$. The physical scale height in units of distance is thus $H \sim R h$. A11 employed an increased scale height close to the sublimation radius ($R_{\rm subl}$) and the wall of the outer disk (at $R_{\rm cav}$) to mimic ``puffed-up'' structures. However, due to the missing angular resolution of the observations, the inner rim structure is not well constrained and there is a degeneracy between the height of the inner rim and the amount of dust in the inner disk ($\delta_{\rm dust}$). Since the main effect of a dusty inner disk is to shield the gap and outer disk from direct stellar irradiation, we only vary $\delta_{\rm dust}$ and do not increase the scale height.

Dust settling is implemented considering two different populations of grains (small $0.005 - 1$ $\mu$m and large $0.005 - 1000$ $\mu$m) following the approach by \cite{dAlessio06}. The scale height for the small population is set to $h$, while the scale height of the larger population to $\chi h$ with $\chi < 1$. The fraction of surface density distributed to the small and large population is $\Sigma_{\rm dust} (1-f)$ and $\Sigma_{\rm dust} f$, respectively. For the outer disk ($R > R_{\rm cav}$), the dust densities thus read 
\begin{eqnarray}
\rho_{\rm dust, small} &=& \frac{(1-f) \, \Sigma_{\rm dust}}{\sqrt{2 \pi} \,  R h} \, \exp\left[-\frac{1}{2} \left( \frac{\pi/2 - \theta}{h} \right)^2 \right] \ \ {\rm and}\\
\rho_{\rm dust, large} &=& \frac{ f \, \Sigma_{\rm dust}}{\sqrt{2 \pi} \, R \chi h} \, \exp\left[-\frac{1}{2} \left( \frac{\pi/2 - \theta}{ \chi h} \right)^2 \right] \ .
\end{eqnarray}

\subsubsection{Dust and PAH opacities} \label{sec:dustpah}

\begin{figure}[htb]
\center
\includegraphics[width=0.9\hsize]{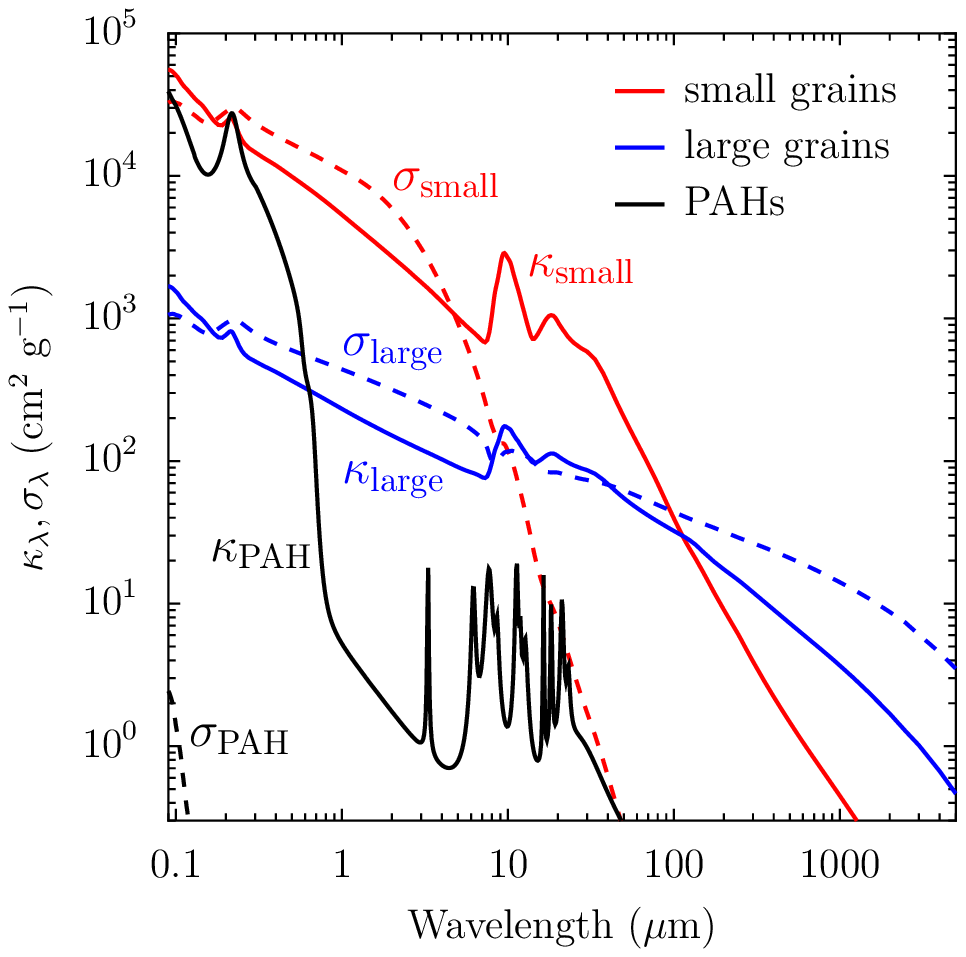}
\caption{Dust and PAH opacities. Absorption ($\kappa$) and scattering ($\sigma$) mass coefficients are given by solid and dashed lines, respectively.}
\label{fig:opac}
\end{figure}

Dust opacities for a standard ISM dust composition following \cite{Weingartner01} are used, consistent with A11. The mass extinction coefficients  are calculated using Mie theory with the \verb!miex! code (\citealt{Wolf04}) and optical constants by \cite{Draine03a} for graphite and \cite{Weingartner01} for silicates. Figure \ref{fig:opac} shows the absorption and scattering opacities for the large grain population (0.005 $\mu$m - 1 mm) and the small grain population (0.005-1 $\mu$m). They are consistent with A11.

The PAH opacities presented in Figure \ref{fig:opac} are given for an ISM abundance and all PAHs assumed to be neutral and in C$_{100}$H$_{25}$. The ISM PAH-to-dust mass ratio is 5 \% (\citealt{Draine07}) corresponding to a C$_{100}$H$_{25}$ abundance of $4 \times 10^{-7}$ relative to hydrogen. The PAH opacities are calculated from \cite{Li01} and \cite{Draine07} using the routines by \cite{Visser07}. In our model, PAHs are mainly important as additional absorbers of UV/optical photons, to heat the gas through the photoelectric effect, and to provide an additional path for H$_2$ formation (through hydrogenated PAHs). For UV/optical wavelengths, the model PAH opacities for neutral and ionized PAHs are the same. As we make no attempt to model the infrared features of PAHs (see \citealt{Visser07}, \citealt{Geers06}), the choice of purely neutral PAHs does not affect our results. We use large PAHs with 100 carbon atoms, since smaller PAHs are predicted to be photodissociated by the strong UV field close to a star (\citealt{Visser07}). We assume that PAHs are well mixed with the gas and their density thus scales with the total gas density. PAHs are absent in regions with FUV radiation stronger than $10^6$ times the interstellar radiation field following the results by \citet{Visser07}. We note that observational evidence of PAHs towards several transition disks exist (\citealt{Brown07}, \citealt{Merin10}) and in one case, PAHs have been imaged to be inside the dust cavity (\citealt{Geers07}).

\subsubsection{Other parameters}

The elemental composition with respect to the total hydrogen abundance is chosen following \cite{Bruderer12}. We assume about one third of the carbon to be volatile, thus not bound in refractory material (Table \ref{tab:param_tdisk}).

The input stellar spectrum is taken to be a black-body of either 6000 or 10000 K, in order to study the effect of different amounts of FUV (6-13.6 eV) photons relative to the bolometric luminosity (as for example in T Tauri or Herbig AeBe stars). The fraction of FUV photons $f_{\rm FUV} = L_{\rm FUV} / L_{\rm bol}$ is $3 \times 10^{-4}$ (for 6000 K) and $7 \times 10^{-2}$ (for 10000 K). The input stellar X-ray spectrum is assumed to be a thermal spectrum of $7 \times 10^7$ K within 1-100 keV. The cosmic-ray ionization rate is taken to be $5 \times 10^{-17}$ s$^{-1}$. The isotropic interstellar radiation field in the UV (\citealt{Draine78,Draine96}) and the cosmic microwave background are also accounted for.

The calculation is carried out on a grid with 100 cells in radial and 80 cells in vertical direction. In vertical direction, the cells are arranged following the scale height at a given radius. In order to properly resolve the inner regions of the inner and outer disk, 50 points are distributed radially in the inner and outer disk each (Figure \ref{fig:repmod_phys}c).

\subsection{Choice of parameters: A grid of models}\label{sec:grid}

Since the goal of our study is to understand the physics and chemistry of the gas in the inner dust hole of transition disks rather than reproducing individual objects, we run a grid of models that covers the range of dust structures found by A11. The assumed parameters are summarized in Table \ref{tab:param_tdisk}.

The distribution of gas is not constrained by A11. We assume the gas profile to follow the total dust profile scaled by a factor. In the outer disk, we set the gas-to-dust ratio to $\Delta_{\rm gas/dust}=100$. Within the gap, the gas is scaled by a constant factor of $\delta_{\rm gas}=10^{-6}$ to 1 relative to the density in the outer disk. This is the principal parameter that is varied.

\begin{figure}
\center
\includegraphics[width=0.9\hsize]{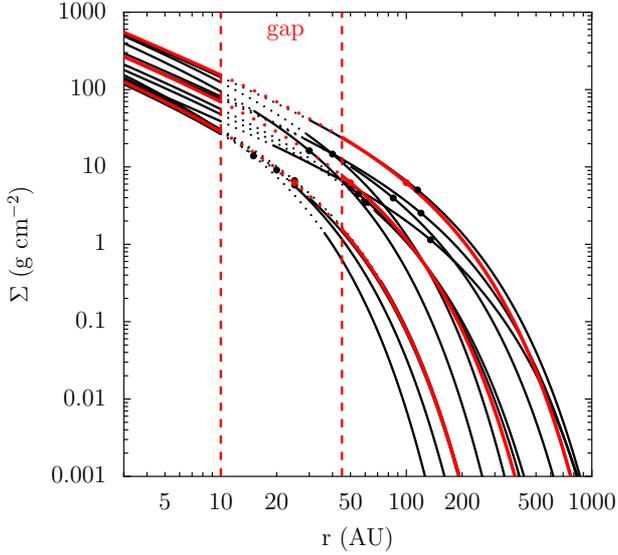}
\caption{Surface density profiles of the observed transition disks (black lines; \citealt{Andrews11}) compared to the model surface density profiles used in this work (red lines). Dotted lines indicate regions within the gap. The red and black dots show $(R_{\rm c}, \Sigma_{\rm c} /e)$.}
\label{fig:allsurfs}
\end{figure}

The surface density power-law index is fixed to $\gamma=1$ following A11. We also fix $\Sigma_{\rm c}$ but vary $R_{\rm c}$ in order to cover the observed surface density profiles (Figure \ref{fig:allsurfs}). 

We run models both with an inner dusty disk present or absent. Therefore, the dust depletion factor $\delta_{\rm dust}$ (Figure \ref{fig:densstruct}) is set to $\delta_{\rm dust}=10^{-5}$ or $\delta_{\rm dust}=10^{-10}$. The higher value is about the median found by A11 and yields an optically thick dusty inner disk at UV to optical wavelengths, which can shield the gap from the stellar radiation. The lower value yields an optically thin dusty inner disk. The remaining amount of dust in these models is negligible and an even lower $\delta_{\rm dust}$ would not change the results. If present, the dusty inner disk extends from the dust sublimation radius to a radius of $R_{\rm gap}=10$ AU. Since the main effect of an inner dusty disk is the shielding of the stellar radiation, which already takes place in the inner wall of the inner disk, models with much smaller gap radii, e.g. $R_{\rm gap}=1$ AU, give very similar results (see Appendix \ref{sec:app_diffgapcav}).

Between a radius of 10 AU and 45 AU, a gap without dust is present. We chose the size of the gap to be relatively large in order to study the physics and chemistry in the gap up to larger radii, which can be probed with current observational facilities. We however note, that the trends found for the 45 AU radius gap also apply to smaller cavity sizes (Appendix \ref{sec:app_diffgapcav}). In the vertical direction, we assume a scale height angle $h_{\rm c}=0.06$ rad and a flaring index $\psi=0.2$ which is the average of A11. The dust-settling parameters $f$ and $\chi$ are fixed to the values by A11.

The luminosity of the star is assumed to be 1, 3 or 10 $L_{\odot}$. The X-ray luminosity is taken to be either $10^{27}$ or $10^{30}$ erg s$^{-1}$. The higher value corresponds to the X-ray luminosities found towards the disks discussed in A11 ($0.5-3 \times 10^{30}$ erg s$^{-1}$). The lower value yields results both in the chemistry and temperature that are close to X-rays being completely absent. 

\begin{table}[tbh]
\caption{Parameters of the transition disk models. Parameters of a representative model are given in bold.}
\label{tab:param_tdisk}
\centering
\begin{tabular}{ll}
\hline\hline
Parameter &  Range \\
\hline
\it{Physical structure} \\
$\gamma$                & 1\tablefootmark{\dagger} \\
$\psi$                  & 0.2 \\
$h_{\rm c}$             & 0.06 rad \\
$\Sigma_{\rm c}$        & 17 g cm$^{-2}$ \\
$R_{\rm c}$             & 25, \textbf{50}, 100 AU \\
$R_{\rm cav}$           & 45 AU\\
$R_{\rm subl}$          & $0.07 \sqrt{L_*/L_\odot}$ AU\tablefootmark{\dagger}\\
$R_{\rm gap}$           & 10 AU\\
$\Delta_{\rm gas-dust}$ & 0.01\tablefootmark{\dagger} \\
$\delta_{\rm gas}$      & $10^{-6}, 10^{-5}, 10^{-4}, 10^{-3}, 10^{-2}, 10^{-1}, 1$\\
$\delta_{\rm dust}$     & $10^{-10}$ (no dusty inner disk), \\ 
                        & $10^{-5}$ (with dusty inner disk) \\
$f$                     & 0.85\tablefootmark{\dagger} \\
$\chi$                  & 0.2\tablefootmark{\dagger} \\
\\
\it{Stellar spectrum} \\
$T_{\rm eff}$           & 6000, \textbf{10000} K\\
$L_{\rm bol}$           & 1, 3, \textbf{10} $L_\odot$\\
$L_{\rm X}$             & $10^{27}$, $\mathbf{10^{30}}$ erg s$^{-1}$ \\
\\
\it{Dust properties} \\
Dust                    & 0.005-1 $\mu$m (small)\\
                        & 0.005-1000 $\mu$m (large)\\
PAHs                    & 0.1, 1.0, \textbf{10} \% w.r.t. ISM abundance\tablefootmark{b}\\
\\
\it{Gas elemental composition\tablefootmark{a}} \\
H                       &      1        \\
He                      &      7.59(-2) \\
C                       &      1.35(-4) \\
N                       &      2.14(-5) \\
O                       &      2.88(-4) \\
Mg                      &      4.17(-6) \\
Si                      &      7.94(-6) \\
S                       &      1.91(-6) \\
Fe                      &      4.27(-6) \\
\hline
\end{tabular}
\tablefoot{\tablefoottext{\dagger}{Fixed, see \cite{Andrews11}.} \tablefoottext{a}{see \cite{Bruderer12}}. a(b) means $a \times 10^b$. \tablefoottext{b}{See Sect. \ref{sec:dustpah}.}}
\end{table}

\subsubsection{Gas and dust masses} \label{sec:masses}

The gas- and dust-masses of the outer disk, the gap and inner disk are summarized in Table \ref{tab:param_mass}. Models with $\delta_{\rm gas} \leq 10^{-1}$ always have the bulk of the total mass in the outer disk rather than the cavity (gap and inner disk). The amount of 1 Jupiter mass ($9.5 \times 10^{-4} M_\odot$) within the gap is reached for $\delta_{\rm gas}=3 \times 10^{-1}, 8 \times 10^{-2}$, and $3 \times 10^{-2}$ (for $R_{\rm c}=$25 AU, 50 AU or 100 AU). One Saturn mass ($2.9 \times 10^{-4} M_\odot$) is accordingly reached for $\delta_{\rm gas}=8 \times 10^{-2}, 2 \times 10^{-2}$, and $9 \times 10^{-3}$.

\begin{table}[tbh]
\caption{Gas- and dust-masses of the different models in $M_\odot$.}
\label{tab:param_mass}
\centering
\begin{tabular}{ll|lll}
\hline\hline
      &                & $R_{\rm c}=25$ AU & $R_{\rm c}=50$ AU & $R_{\rm c}=100$ AU \\
\hline                      
$M_{\rm dust}$ & outer disk  & 1.2(-5)           & 1.2(-4)           & 7.7(-4)            \\
& gap            & 0                 & 0                 & 0 \\
 & inner disk   & 2.4(-5)$\delta_{\rm dust}$  & 5.3(-5)$\delta_{\rm dust}$  & 1.1(-4)$\delta_{\rm dust}$ \\
\hline
$M_{\rm gas}$ & outer disk   & 1.2(-3)           & 1.2(-2)           & 7.7(-2)            \\
 & gap         & 3.8(-3)$\delta_{\rm gas}$ & 1.2(-2)$\delta_{\rm gas}$ & 3.2(-2)$\delta_{\rm gas}$ \\
 & inner disk  & 2.4(-3)$\delta_{\rm gas}$ & 5.3(-3)$\delta_{\rm gas}$ & 1.1(-2)$\delta_{\rm gas}$ \\
\hline
\end{tabular}
\tablefoot{a(b) means $a \times 10^b$.}
\end{table}

\subsection{Caveats}

A number of caveats apply to the models presented here: As discussed in \cite{Bruderer12}, the calculated gas temperature is prone to large uncertainty. A comparison of PDR models run for a density/FUV intensity combination corresponding to the outer disk has revealed a scatter of a factor of a few in gas temperature (\citealt{Roellig07}). High-temperature tracers such as ro-vibrational lines (e.g. of CO) or high-$J$ CO lines are  more affected by this uncertainty than lower temperature lines (e.g. low-$J$ lines of CO) which form deeper in the disk.

From the comparison of different PDR models and the uncertainty analysis of the chemical network (\citealt{Vasyunin08}), we estimate the typical uncertainty of the low-$J$ CO line intensities modeled in this work to be about a factor of two. Since the dynamic range of CO intensities considered here is several orders of magnitude, these uncertainties do not affect our conclusions.

The analytical density structure of our models is a simplification and does not account for the hydrostatic equilibrium of the gas or dust temperature. One of the main effects of the hydrostatic equilibrium is a ``puffed-up'' inner rim. However, it is unclear at what point such a structure would become unstable and being blown away as a wind. Furthermore, \citet{Min09} find that the calculation of the vertical structure can lead to density waves propagating on the surface of the disk. In order to resolve these two issues, the structure calculation should be coupled with a hydrodynamical simulation which is beyond the scope of this study and computationally very challenging. Also, the physical mechanism leading to the formation of the dust hole needs to be understood in order to run a more complete simulation.

The dust size distribution and composition is not well constrained from observations and the assumed power-law distributions are only an approximation. In future studies, we plan to incorporate the results of dust evolutionary models (\citealt{Birnstiel10}).

Our models are solved for steady-state conditions. While the chemical time-scales are usually short, mixing in vertical or radial direction could smooth out transitions (e.g. \citealt{Ilgner06b}, \citealt{Heinzeller11}, \citealt{Semenov11}). Abundances of CO isotopologues are approximated by scaling the abundance of the main isotopologue. The CO and H$_2$ photodissociation rate and the C ionization rate are calculated with self- and mutual-shielding factors in a 1+1D way. The accuracy of this treatment has been verified in \cite{vanZadelhoff03} and \cite{Visser09b}.

%
%
\section{Physical structure} \label{sec:rep_physstructure}

In this section, we will discuss the physical structure (density, gas/dust-temperature, \ldots) of one representative model. For the representative model, we choose $R_{\rm c}=50$ AU, $T_{\rm eff}=10000$ K, $L_{\rm bol}=10$ $L_\odot$, and a X-ray luminosity of $L_{\rm X}=10^{30}$ erg s$^{-1}$ (Table \ref{tab:param_tdisk}).

\begin{figure*}[!htb]
\includegraphics[width=\hsize]{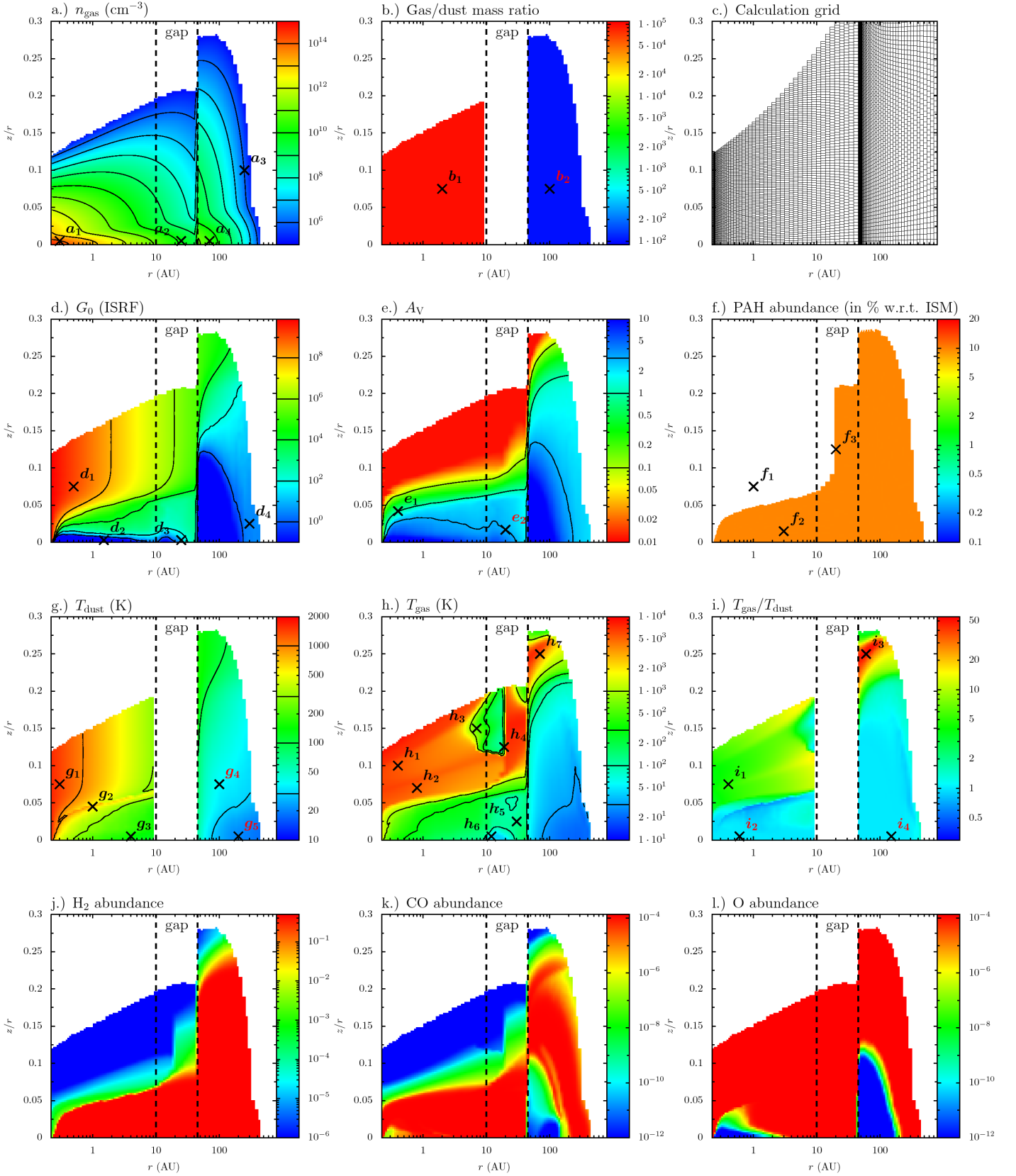}
\caption{Physical conditions in a representative model (Table \ref{tab:param_tdisk}) with $\delta_{\rm gas}=10^{-2}$ and $\delta_{\rm dust}=10^{-5}$ (dusty inner disk present). Only regions with a gas density larger than $10^5$ cm$^{-3}$ are shown. The contour lines on the panels match to the lines on the color bars. The panels provide \textbf{a)} Gas density structure, \textbf{b)} Gas/dust mass ratio, \textbf{c)} Calculation grid, \textbf{d)} FUV radiation field, \textbf{e)} Visual extinction, \textbf{f)} Abundance of PAHs, \textbf{g)} Dust temperature, \textbf{h)} Gas temperature, \textbf{i)} Ratio of gas to dust temperature, \textbf{j)} H$_2$ abundance, \textbf{k)} CO abundance, and \textbf{l)} O abundance.}
\label{fig:repmod_phys}
\end{figure*}

Figure \ref{fig:repmod_phys} shows the physical structure of the representative model with $\delta_{\rm gas} = 10^{-2}$ and a dusty inner disk present ($\delta_{\rm dust}= 10^{-5}$). The different subfigures are given in the $(\log(r),r/z)$-space, so that horizontal lines correspond to a constant height ($r/z$) and thus to direct rays from a position in the disk to the star. We apply a density cut-off at $10^5$ cm$^{-3}$, since the column density above it is small and does not contribute to the emission.

The density given in Figure \ref{fig:repmod_phys}a shows a sharp drop at the cavity radius of 45 AU, due to the scaling of the inner disk density with $\delta_{\rm gas}$. The density in the midplane of the inner disk at $r < 1$ AU is up to $10^{14}$ cm$^{-3}$ even for $\delta_{\rm gas}=10^{-2}$ (label $a_1$). In the gap, the midplane density reaches $10^{9} - 10^{11}$ cm$^{-3}$ (label $a_2$). In the outer disk, the density cuts off to larger radii due to the exponential decrease of the surface density profile (label $a_3$). In the midplane of the outer disk, densities of $\sim 10^{10}$ cm$^{-3}$ are reached. The gas-to-dust ratio in the inner disk (Figure \ref{fig:repmod_phys}b, label $b_1$) is with $10^5$ a factor of $\delta_{\rm dust}/\delta_{\rm gas}=1000$ larger compared to the outer disk (100, label $b_2$). The gas-to-dust ratio is not defined in the gap, due to the absence of dust.

The strength of the UV field in units of the interstellar radiation field $G_0$ is given in Figure \ref{fig:repmod_phys}d. Here, $G_0=1$ refers to the interstellar radiation field defined as in \cite{Draine78} $\sim 2.7 \times 10^{-3}$ erg s$^{-1}$ cm$^{-2}$ with photon-energies $E_\gamma$ between 6 eV and 13.6 eV. In the inner, upper disk, very high UV strengths of $G_0 \gtrsim 10^9$ are reached (label $d_1$), while the inner midplane is shielded from UV radiation ($G_0 < 1$, label $d_2$). In the gap, both shielding by the dusty inner disk and the PAH opacity contribute to the attenuation of the FUV radiation (label $d_3$). In the outer disk, also external radiation from the interstellar radiation field contributes to the FUV field (label $d_4$). 

The UV extinction, measured in $A_V$, is given in Figure \ref{fig:repmod_phys}f. Following BR12, we calculate the extinction from the ratio of the attenuated to the only geometrically diluted flux, $\tau_{\rm FUV} = - \ln(F_{\rm attenuated}/F_{\rm unattenuated})$, where $F$ is integrated over the FUV spectrum. In the inner disk, $A_V=1$ is reached even in the innermost part (label $e_1$). The decrease of height with distance of the $A_V=3$ surface inside the gap (label $e_2$) indicates that some of the UV radiation is scattered by the small dust of the inner disk into the gap. PAHs do not contribute to the scattering, as their scattering-cross section is small (Figure \ref{fig:opac}).

The abundance of PAHs with respect to the ISM abundance is given in Figure \ref{fig:repmod_phys}f. In the inner, upper disk, PAHs are photodissociated due to the intense UV radiation from the star (label $f_1$). The small dust in the dusty inner disk is able to shield the UV radiation sufficiently to have PAHs surviving (label $f_2$). In absence of a dusty inner disk, PAHs could only survive at radii $\gtrsim 20$ AU (label $f_3$), where geometrical dilution decreases the intensity of the stellar UV radiation sufficiently. The presence of a dusty inner disk has thus influence on the gas in the gap, as it allows PAHs to survive to smaller radii. PAHs affect the gas in the gap through FUV attenuation, the chemistry (H$_2$ formation on PAHs) and gas heating through the photoelectric effect.

The dust temperature $T_{\rm dust}$ is presented in Figure \ref{fig:repmod_phys}g. It is only defined in the inner and the outer disk, where dust is present. In the inner upper disk (label $g_1$), the dust heats up to the sublimation temperature of $\sim 1500$ K. At the edge of the region where PAHs are present, the additional opacity in the UV leads to additional dust heating, which is seen in the inner disk (label $g_2$). In the midplane of the inner disk, dust temperatures between 100 K and 300 K are found (label $g_3$). The bulk of the outer disk is at dust temperatures between 30 and 100 K (label $g_4$). Dust temperatures are below 30 K only in the outer disk, close to the midplane (label $g_5$).

\subsection{Gas temperature} \label{sec:gastemp}

The gas temperature $T_{\rm gas}$ and the ratio of gas to dust temperature ($T_{\rm gas}$/$T_{\rm dust}$) are presented in Figure  \ref{fig:repmod_phys}h and \ref{fig:repmod_phys}i, respectively. The latter ratio is only defined in regions where the dust temperature is defined as well. 

In the upper atmosphere of the inner disk, gas temperature up to several 1000 K are reached (label $h_1$). In this part of the disk, the main heating mechanism is either coulomb heating through X-rays or collisional de-excitation of \ion{Fe}{II} which is pumped by optical/UV stellar radiation. This back-heating by \ion{Fe}{II} has also been found to be an important heating agent in other models (see \citealt{Woitke09}). The main cooling in this region is either through \ion{O}{I} or Ly$\alpha$ emission. Moving deeper into the disk (label $h_2$), the H$_2$ abundance is considerable and collisional de-excitation of UV pumped H$_2$ is an important heating mechanism. 

Moving to the upper atmosphere at larger radii $\gtrsim 10$ AU, both optical/UV back-heating by atoms and X-ray heating get less effective (label $h_3$) and the gas temperature decreases. Further out, the gas temperature in the upper atmosphere increases again at the radius where PAHs can survive (label $h_4$). Since the PAHs also increase the H$_2$ abundance, through formation of H$_2$ on their surface, heating by UV pumped H$_2$ becomes important. Both at $h_3$ and $h_4$, atomic lines (\ion{O}{I}, \ion{C}{II}, \ion{Si}{II}) dominate the cooling. The upper atmosphere of the outer disk (label $h_7$) is heated to several 1000 K by photoelectric heating on PAHs and small grains and also UV pumped H$_2$.

In the midplane of the disk, the gas temperature is generally lower than in the upper atmosphere. In the midplane of the inner and outer disk, $T_{\rm gas}$/$T_{\rm dust}$ is close to 1 (labels $i_2$, $i_4$). For comparison, $T_{\rm gas}$/$T_{\rm dust}$ in the upper atmosphere is a few in the inner disk (label $i_1$) and $\sim 50$ in the outer disk (labels $i_3$). In the midplane of the inner disk, at distances $R < 1$ AU, molecules (CO, H$_2$O) act as heating agents through the pumping by the strong IR radiation field. In this region gas-grain cooling is important despite the high gas-to-dust ratio. Further out ($r \gtrsim 2$ AU), the IR radiation field is weaker and the molecules cool, while gas-grain collisions heat the gas. Cosmic-ray heating then also becomes important. In the midplane of the gap (label $h_5$, $h_6$), heating by PAHs and small grains is balanced with molecular and atomic cooling (CO, $^{13}$CO and [\ion{O}{I}]) . The gas temperature in the gap is still mostly above 100 K, except in the shadow behind the inner disk (label $h_6$). In the gap, dust is absent and neither gas-grain-heating nor -cooling takes part. In the midplane of the outer disk, the gas-grain collisions are again present and gas and dust are well coupled ($T_{\rm gas}/T_{\rm dust} \sim 1$, label $i_4$). 

%
%
\section{Survival of molecules in the cavity} \label{sec:survival}

The abundance of H$_2$ and CO for three models with different amounts of gas in the cavity is shown in Figure \ref{fig:inner_H2CO}. The models have $\delta_{\rm gas} = 1, 10^{-4}$, and $10^{-6}$ and a dusty inner disk present. All other parameters are set to the values of the reference model (Table \ref{tab:param_tdisk}). 

\subsection*{H$_2$}

\begin{figure*}[!htb]
\includegraphics[width=\hsize]{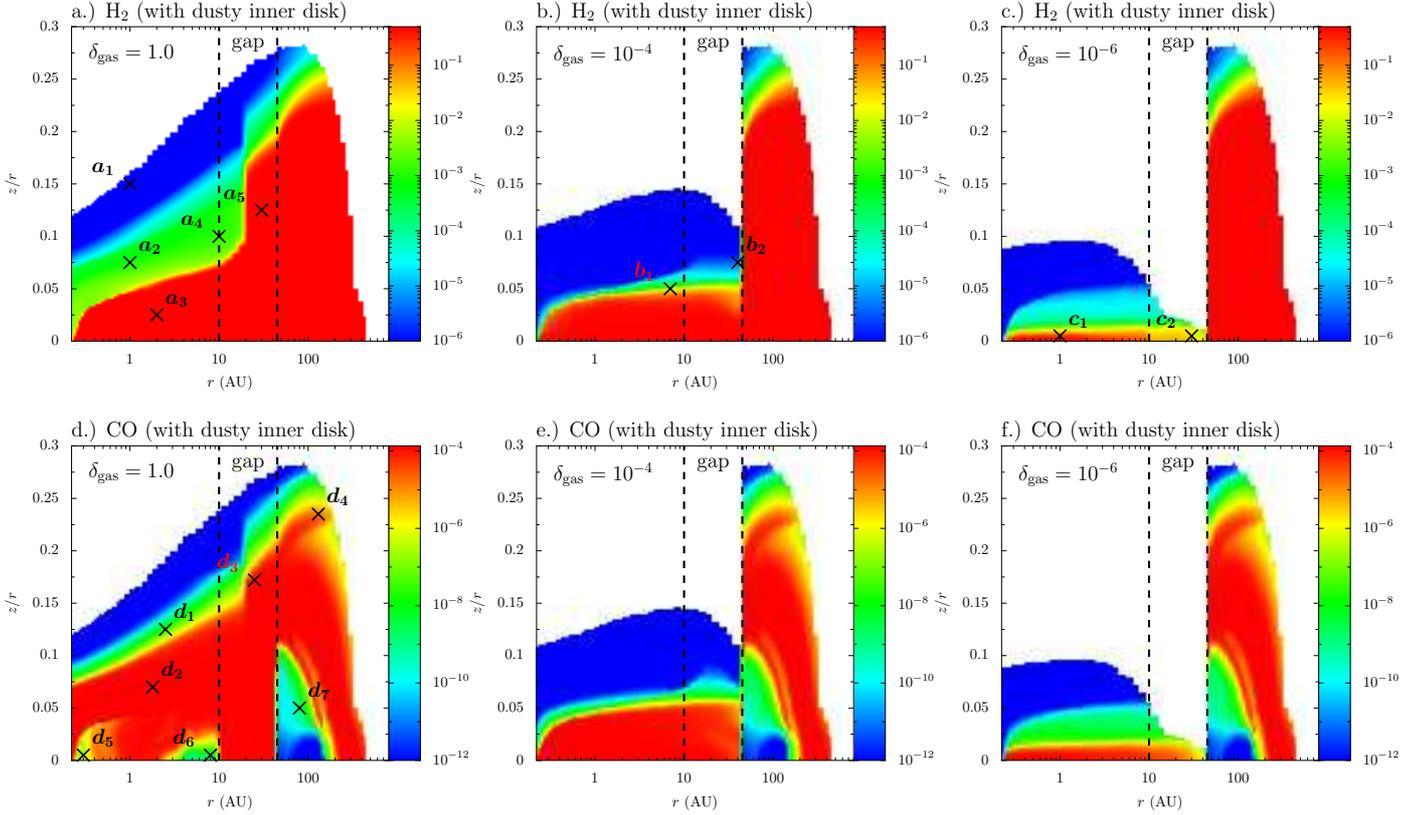}
\caption{H$_2$ and CO abundance in the representative model (Table \ref{tab:param_tdisk}) with different gas content in the cavity ($\delta_{\rm gas} = 1, 10^{-4}$ and $10^{-6}$) and a dusty inner disk present ($\delta_{\rm dust}=10^{-5}$). Only regions with gas density larger than $10^5$ cm$^{-3}$ are shown. \textbf{Upper panels:} CO abundance. \textbf{Lower panels:} H$_2$ abundance.}\label{fig:inner_H2CO}
\end{figure*}

Molecular hydrogen (H$_2$) in the model with $\delta_{\rm gas}=1$ is photodissociated to fractional abundances below $10^{-6}$ in the upper atmosphere of the inner disk (label $a_1$ in Figure \ref{fig:inner_H2CO}). Going deeper into the disk, self-shielding reduces the photodissociation and below a plateau with fractional abundance of $10^{-4} - 10^{-2}$ (label $a_2$), hydrogen is fully in molecular form (label $a_3$). The edge of the dust free gap at $r=10$ AU is not seen in the H$_2$ abundance (label $a_4$). This is because H$_2$ can not only form on dust grains, but also through the reaction ${\rm H}^- + {\rm H} \rightarrow {\rm H}_2 + e^-$, if this reaction can compete with the  (quick) photodetachment of H$^-$. H$^-$ is produced by electron attachment of atomic hydrogen. Outside the radius of PAH photodestruction (label $a_5$), hydrogen is fully molecular, because of H$_2$ formation on PAHs. The outer disk is molecular up to large heights, because H$_2$ forms efficiently on dust grains present in the outer disk. We conclude, that for the formation of H$_2$, and to start the chemistry of other species (e.g. CO), either dust or PAHs need to be present to form H$_2$ on the surface or conditions need to allow H$_2$ formation through ${\rm H} + {\rm H}^-$.

Under what conditions is H$_2$ formation through H$^-$ efficient? Since photodetachment of H$^-$ is very rapid, the reaction of H$^-$ with H needs to compete with it. The photodetachment rate of H$^-$ is $\sim 2 \times 10^{-7} G_0 n_{{\rm H}^-}$ cm$^{-3}$ s$^{-1}$ for an unshielded 10000 K blackbody spectrum. The reaction of H$^-$ with H on the other hand runs at $1.3 \times 10^{-9} n_{\rm H} n_{{\rm H}^-}$ cm$^{-3}$ s$^{-1}$ (\citealt{Woodall07}). At the inner edge of the gap ($r=10$ AU), the midplane density of the representative model is  $\sim 10^{13} \delta_{\rm gas}$ cm$^{-3}$ and the unshielded FUV field corresponds to $G_0 \sim 3 \times 10^6$. Thus, for $\delta_{\rm gas} > 5 \times 10^{-5}$, the density is high enough for H$_2$ production to be more important than photodetachment. Is the overall rate of H$_2$ formation also considerable? The electron-attachement rate of  hydrogen (${\rm H} + e^- \rightarrow {\rm H}^- + \gamma$) is low, $\sim2.8 \times 10^{-17} \sqrt{T} n_{\rm H} n_{e^-}$. Assuming all carbon being photoionized ($n_{e^-} \sim 10^{-4} n_{\rm H}$) we can estimate the H$_2$ formation rate under the assumption that all H$^-$ is eventually turned into H$_2$. The H$_2$ formation rate then is $\sim 1.3 \times 10^{-9} n_{\rm H} n_{{\rm H}^-} \sim 2.9 \times 10^{-21} \sqrt{T} n_{\rm H}^2$ cm$^{-3}$ s$^{-1}$. For comparison the H$_2$ formation rate on dust grains is $\sim 3 \times 10^{-18} \sqrt{T} n_{\rm H} n_{\rm tot}$ cm$^{-3}$ s$^{-1}$, assuming sticking/accommodation coefficients of 1 (an upper limit) and a gas-to-dust ratio of 100. Formation on dust grains is thus a factor of 1000 more efficient. However, even in this model with dusty inner disk present, the gas-to-dust ratio in the inner disk can be much higher than 100 with values of $10^7$ reached (for $\delta_{\rm gas}=1$ and $\delta_{\rm dust}=10^{-5}$) and H$_2$ formation through H$^-$ more efficient than on the remaining small amount of dust. We conclude that H$_2$ formation through H$^-$ can become more important than formation on dust under the special conditions of high density, high electron abundance, and a high gas-to-dust ratio.

We now study the effect of lowering the gas mass in the cavity (decreasing $\delta_{\rm gas}$). Lowering $\delta_{\rm gas}$ from $1$ has different effects on our model. First, the gas density at a given point inside the cavity is scaled down by a factor $\delta_{\rm gas}$. Second, since we assume PAHs to be mixed with the gas, the PAH density is also lowered so the FUV radiation can penetrate deeper into the disk. The importance of PAHs to the FUV shielding is understood by their contribution to the opacity shown in Figure \ref{fig:opac}. In the models discussed here, the PAH abundance is a factor of 10 lower, but the gas-to-dust ratio a factor of $10^5 \delta_{\rm gas}$ higher, so the PAHs dominate the FUV opacity in the dusty inner disk for $\delta_{\rm gas} \gtrsim 10^{-4}$. 

The effect of stronger local FUV irradiation due to less shielding by PAHs is seen in the model with $\delta_{\rm gas}=10^{-4}$. The H/H$_2$ transition moves down (label $b_1$) and due to the reduced density, the region outside the PAH photodestruction radius extending to larger height (label $a_5$) is not found in this model (label $b_2$). The steeper gradient of the H/H$_2$ transition in this model is explained by a thermo-chemical feedback loop: Models with lower $\delta_{\rm gas}$ form H$_2$ deeper in the disk where less FUV radiation penetrates. The heating is thus less efficient and the gas temperature lower at the H/H$_2$ transition. Conversely, models with higher $\delta_{\rm gas}$ have higher temperatures at the H/H$_2$ transition and the H$_2$ formation rate is suppressed by the higher  temperature.

The effect of a decreased density on the H$_2$ abundance can be roughly understood by the fact that gas formation rates scale as $n^2$, while photodissociation rates only scale with the density $n$. Assuming a simplistic H/H$_2$ chemistry, only considering H$_2$ formation with rate coefficient $F$ and H$_2$ photodissociation with rate coefficient $D G_0$, the rate equations for the H$_2$ abundance reads $\dot{n}_{{\rm H}_2}= F n \left( n - 2 n_{{\rm H}_2} \right) - D G_0 n_{{\rm H}_2}$. Defining $\alpha=n F / (D G_0)$, the H$_2$ density in steady state is 
\begin{equation}
n_{{\rm H}_2} = n \frac{ \alpha}{1 + 2 \alpha} 
\end{equation}
Consequently, the fractional abundance of H$_2$ ($n_{{\rm H}_2}/n$) scales about linearly with $\alpha \propto n / G_0$ until it settles off at $1/2$ for high values of $\alpha$.

Decreasing $\delta_{\rm gas}$ further to $\delta_{\rm gas} = 10^{-6}$ results in the presence of H$_2$ within the cavity only in a geometrically thin layer close to midplane (label $c_1$). In the gap (label $c_2$), hydrogen is not in full molecular form anymore, but still reaches appreciable fractional abundances of $\sim 10^{-2} - 0.1$.

In the outer disk, the amount of H$_2$ does not differ between the three models with different $\delta_{\rm gas}$, because the upper layers of the outer disk, where H$_2$ photodissociation proceeds, is posed to direct irradiation by the star independent of $\delta_{\rm gas}$. Lowering $\delta_{\rm gas}$  however allows direct irradiation of the outer disk wall to regions closer to the midplane. Thus, photodissociation can proceed along the wall.

\subsection*{CO}

The chemistry of carbon monoxide (CO) is more complex compared to H$_2$. In this section we will concentrate on CO, the chemistry of C and C$^+$ will be discussed in more detail in Section \ref{sec:atomcarbon}. First looking at the model with $\delta_{\rm gas}=1$, we find CO like H$_2$ to be photodissociated in the upper atmosphere (label $d_1$). In this region, carbon is mainly in the form of C$^+$. Moving deeper into the disk, a thin layer of neutral carbon connects to the zone where carbon is fully bound in CO (label $d_2$). For this particular model, the C$^+$/C/CO transition is at larger heights in the atmosphere compared to the H/H$_2$ transitions. This is because, contrary to H$_2$, CO is formed more quickly at higher temperatures through CH$^+$. As soon as a certain level of H$_2$ is available in a high temperature region, C$^+$ reacts with H$_2$ to CH$^+$. This reaction has a high energy barrier (4640 K), but hot (vibrationally excited) H$_2^*$ can help to overcome this barrier. The CH$^+$ is then converted by reactions with O to CO$^+$ and with H$_2$ to HCO$^+$ which recombines to CO (\citealt{Jonkheid07}). Moving outward to larger radii, the CO abundance increases at the edge of the PAH destruction radius (label $d_3$), because additional H$_2$ (label $a_5$) leads to more efficient CO formation and a decreased photodissociation rate, due to mutual CO-H$_2$ shielding. In the outer, upper disk, a ``warm finger'' (label $d_4$) of CO is found, where CO formation is again initiated by CH$^+$ (see e.g. BR12).

Close to the midplane of the model with $\delta_{\rm gas}=1$, carbon is not always in the form of CO, unlike hydrogen in H$_2$. In the inner disk (label $d_5$), the reaction of O with H$_2$ to OH followed by OH $+$ H$_2$ to water drives all oxygen into water. Both reactions have considerable activation barriers and only proceed at temperatures $\gtrsim 250$ K. Since the midplane is well shielded from UV radiation, photodissociation of OH and water are not efficient. In this region, carbon is bound into methane (CH$_4$), due to the absence of free oxygen to form CO. Further out in the midplane, the temperature is insufficient for OH and H$_2$O formation and carbon is in CO. In the outer part of the dusty inner disk (label $d_6$), the density/temperature combination allows freeze out of water. Thus, oxygen is locked-up in ices and carbon is again bound in CH$_4$. In the midplane of the gap, no dust for freeze out is available and carbon remains in CO. Moving to the outer disk (label $d_7$), dust is again present and binds oxygen into the ice and the CO abundance in the midplane is low even though dust temperatures are too high for CO freeze-out.

Lower amounts of gas in the cavity ($\delta_{\rm gas} < 1$) affect CO in a similar way as H$_2$. The C$^+$/CO transition shifts to lower heights in the disk. The model with $\delta_{\rm gas}=10^{-4}$ does not form additional CO in the upper disk at the PAH destruction radius (label $d_3$), because no additional H$_2$ is found at that position (label $b_2$). In the midplane of this model, the regions with decreased CO abundance (labels $d_5$ and $d_6$) are absent. This is because the lower density and subsequently lower shielding by PAHs leads to less efficient formation of OH/H$_2$O and more photodissociation in the inner part of the inner disk (label $d_5$). In the outer part of the inner disk, freeze out is less efficient and photodesorption/photodissociation runs more quickly (label $d_6$). Thus, oxygen is not locked in water or water ice anymore and carbon remains in CO. By decreasing $\delta_{\rm gas}$, the outer water ice reservoir first disappears before the inner gas phase water vanishes (see Figure \ref{fig:repmod_phys}k with $\delta_{\rm gas}=10^{-2}$).

For $\delta_{\rm gas}=10^{-6}$, CO is still present within the cavity, but the the C$^+$/CO transition has moved further down. In the gap, the abundance of CO is now slightly below the abundance of volatile carbon, i.e. not all carbon is bound in CO. The outer disk does not show differences in the CO abundance between models with different $\delta_{\rm gas}$, for the same reasons as discussed for H$_2$ in the previous paragraph. 

In summary, we find that H$_2$ and CO can still survive for $\delta_{\rm gas}=10^{-6}$, corresponding to only 0.004 M$_{\rm Earth}$ inside the gap, because they are both very robust against photodissociation, through self- and mutual-shielding. This low value can be better understood by turning it into a column density. Distributing 0.004 M$_{\rm Earth}$ homogeneously over the gap ($r=10-45$ AU) yields a total column density of $3 \times 10^{19}$ cm$^{-2}$. For H$_2$ or CO self-shielding to become active, a column of order $10^{15}$ cm$^{-2}$ is sufficient (\citealt{Draine96}, \citealt{Visser09b}). In addition, the dust in the inner disk provides an attenuation $A_V \gtrsim 5$ and is thus able to shield direct FUV radiation even if PAHs are absent. Thus, the survival of H$_2$ and CO even for this low amount of gas in the gap seems plausible. The presence of the molecules does however not necessarily mean that their emission is strong enough for a detection (see Sect. \ref{sec:molemission})

\subsection{The effect of a dusty inner disk on molecular gas} \label{sec:innerdisk}

\begin{figure*}[!htb]
\sidecaption
\includegraphics[width=0.66\hsize]{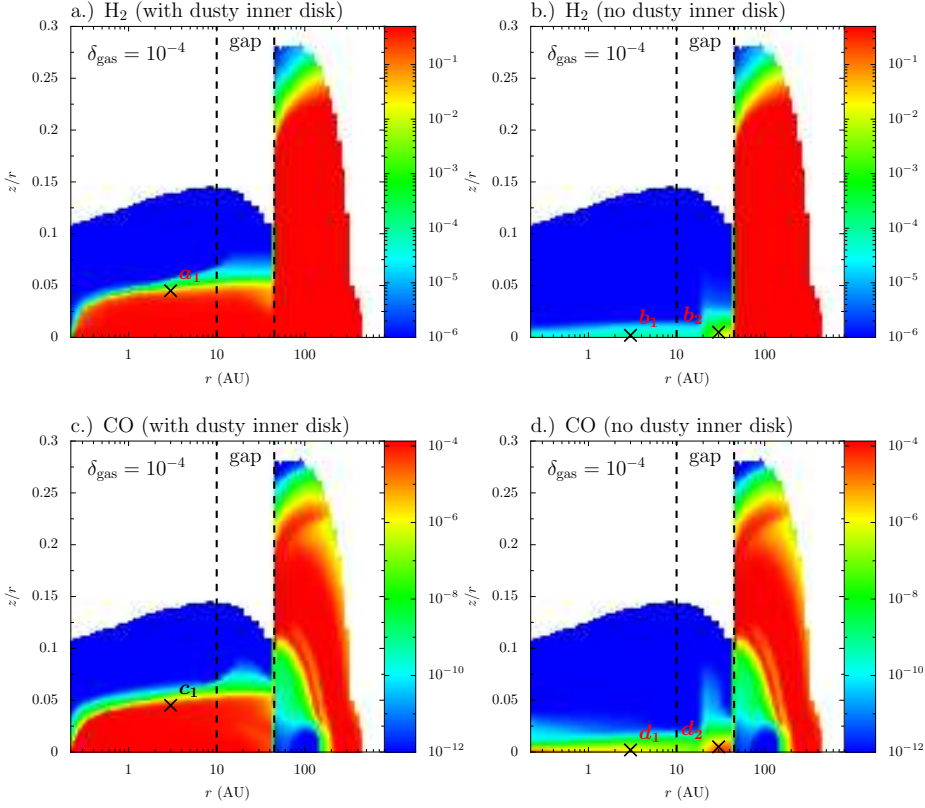}
\caption{H$_2$ and CO abundance of a representative model (Table \ref{tab:param_tdisk}) with dusty inner disk ($\delta_{\rm dust}=10^{-5}$) compared to a model without dusty inner disk ($\delta_{\rm dust}=10^{-10}$). Both models have $\delta_{\rm gas}=10^{-4}$. \textbf{Upper panels:} H$_2$ abundance. \textbf{Lower panels:} CO abundance.} \label{fig:noinner_H2CO}
\end{figure*}

The previous sections have shown that the attenuation of stellar FUV radiation is key for the survival of molecules in the cavity. This raises the question if molecules can still survive, if no FUV shielding by a dusty inner disk is provided. In Figure \ref{fig:noinner_H2CO}, the abundance of H$_2$ and CO of a model with an optically thick dusty inner disk ($\delta_{\rm dust}=10^{-5}$) is compared to one without ($\delta_{\rm dust}=10^{-10}$). Note that there is still an inner gas disk in the model without dusty inner disk (Figure \ref{fig:densstruct}). The representative model (Table \ref{tab:param_tdisk}) with $\delta_{\rm gas}=10^{-4}$ is shown. 

The differences in the abundance structure of H$_2$ and CO between the model with and without dusty inner disk are evident. The model without dusty inner disk has H$_2$ and CO present only in a geometrically thin layer close to the midplane (labels $b_1$ and $d_1$), while the model with dusty inner disk present has a H/H$_2$ and C$^+$/CO transition at considerable height in the disk (labels $a_1$ and $c_1$). The lower abundance of CO and H$_2$ in the model without dusty inner disk are explained by the missing FUV absorption of the dust. This leads to stronger FUV irradiation of the gap and subsequently, PAHs can only survive outside the radius where geometrical dilution decreases the stellar radiation sufficiently. This is around $r\sim 20$ AU and can be seen by an enhanced H$_2$ and CO abundance (labels $b_2$ and $d_2$). The midplane gas temperature at radii $<20$ AU, where no PAHs survive, is nevertheless high (several 1000 K). This is because no gas-grain coupling can cool the gas, while the  strong continuum radiation field turns lines into heating agents (Sect. \ref{sec:gastemp}). The warm temperatures explain why a thin layer of gas close to the midplane forms CO at abundances $\sim 10^{-4}$, while the H$_2$ abundance does not exceed $10^{-4}$. The remaining H$_2$ at radii $<20$ AU is formed through ${\rm H}^- + {\rm H} \rightarrow {\rm H}_2 + e^-$ (Sect. \ref{sec:survival}), because neither dust grains nor PAHs are present.

Models with more gas in the cavity ($\delta_{\rm gas} > 10^{-4}$) and no dusty inner disk present have more gas in H$_2$ and CO (not depicted). For $\delta_{\rm gas}=1$, the C$^+$/CO transition is at similar heights compared to the model with a dusty inner disk present, because of ``warm'' CO formation through ${\rm C}^+ + {\rm H}_2$. The two regions where carbon is in the form of methane (Figure \ref{fig:inner_H2CO}d, label $d_5$ and $d_6$) do however not exist because of the missing FUV attenuation. The H$_2$ abundance of this model is also similar to the model with dusty inner disk. However, within a few AU, the FUV radiation is too strong for formation via H$^-$ and the H$_2$ abundance is decreased. This is a result of the surface density profile $\Sigma \sim r^{-1}$ while the geometrical dilution of FUV radiation $\sim r^{-2}$. 

The presence of an optically thick dusty inner disk which shields the gas from direct stellar irradiation is thus crucial for molecules to survive in the gap. In absence of dust and PAHs, other chemical mechanisms like the formation of H$_2$ through H$^-$ can still start a chemistry and lead to the presence of molecules. Key for the existence of molecules is a way for the formation of H$_2$. Once an appreciable level of H$_2$ (fractional abundances $\gtrsim 10^{-4}$) is reached, CO can also form.

\subsection{Atomic carbon} \label{sec:atomcarbon}

\begin{figure*}[!bth]
\includegraphics[width=\hsize]{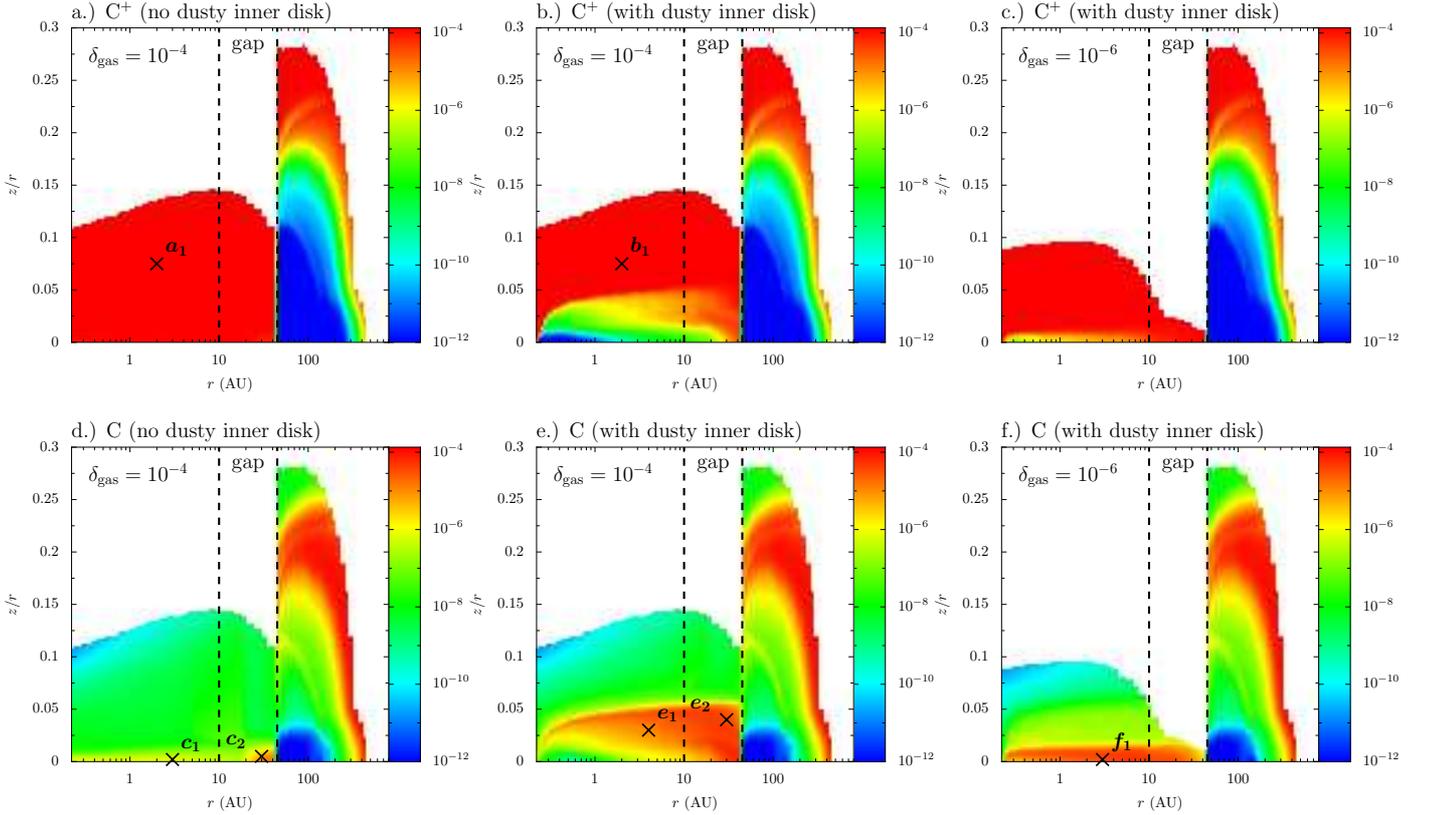}
\caption{C and C$^+$ abundance of the representative models (Table \ref{tab:param_tdisk}) with $\delta_{\rm gas}=10^{-6}$ and no dusty inner disk ($\delta_{\rm dust}=10^{-10}$) or $\delta_{\rm gas}=10^{-4}, 10^{-6}$ with dusty inner disk ($\delta_{\rm dust}=10^{-5}$). \textbf{Upper panels:} C$^+$ abundance. \textbf{Lower panels:} C abundance.}
\label{fig:rep_carbon}
\end{figure*}

Figure \ref{fig:rep_carbon} shows the abundance of C and C$^+$ in models with $\delta_{\rm gas}=10^{-6}$ (dusty inner disk present) and $\delta_{\rm gas}=10^{-4}, 10^{-6}$ (no dusty inner disk present). The other parameters are set to to the reference  model (Table \ref{tab:param_tdisk}).

In the model without dusty inner disk and $\delta_{\rm gas}=10^{-4}$, carbon in the gap is mostly atomic and ionized (label $a_1$). Neutral carbon is only abundant close to the midplane with a fractional abundance $\sim 10^{-6}$ (label $c_1$). There, also some CO is present (Figure \ref{fig:noinner_H2CO}, label $d_1$) and C is produced by CO photodissociation. Additional C is produced by the photodissociation of CH or the reaction ${\rm CH} + {\rm H } \rightarrow {\rm C} + {\rm H}_2$. CH originates from the recombination or photodissociation of CH$_2$, CH$_2^+$, or CH$_3^+$. These molecules form starting with the reaction of C$^+$ with H$_2$ to CH$^+$. Outside the radius of PAH photodestruction (label $c_2$), neutral carbon is more abundant, because of FUV shielding by PAHs and charge exchange with ionized carbon (${\rm PAH}^0 + {\rm C}^+ \rightarrow {\rm PAH}^+ + {\rm C}$, see BR12). The abundance structure and chemistry of carbon in the outer disk is similar to BR12 and will not be discussed here.

The model with dusty inner disk present and $\delta_{\rm gas}=10^{-4}$ also has carbon in the upper atmosphere in C$^+$. Below this zone, neutral carbon C reaches high abundances of up to $\sim 10^{-4}$ in a layer which gets wider to larger distances (label $e_1$). It extends down to the midplane in the gap (label $e_2$). While C is mainly produced by CO and CH photodissociation at heights in the disk, also charge exchange of C$^+$ with PAHs becomes important deeper in the atmosphere. Decreasing the amount of gas in the cavity (smaller $\delta_{\rm gas}$) shifts the C$^+$/CO transition down. Subsequently, also the C layer is found closer to the midplane. The model with dusty inner disk and $\delta_{\rm gas}=10^{-6}$ only has a thin layer of neutral carbon C close to the midplane. 

%
%

\section{Molecular line emission} \label{sec:molemission}

In this section, the molecular line emission of pure rotational lines of CO, observable by ALMA, is discussed. We will first show vertical column densities, then study the line formation including opacity effects and finally synthetic images. Equations to convert between different intensity units are provided in Appendix \ref{sec:app_fluxconv}. The analysis of the atomic fine structure lines ([\ion{C}{I}], [\ion{C}{II}], and [\ion{O}{I}]) are presented in Sect. \ref{sec:herschel}.

\subsection{Column densities} \label{sec:column}

\begin{figure*}[!hbt]
\sidecaption
\includegraphics[width=0.66\hsize]{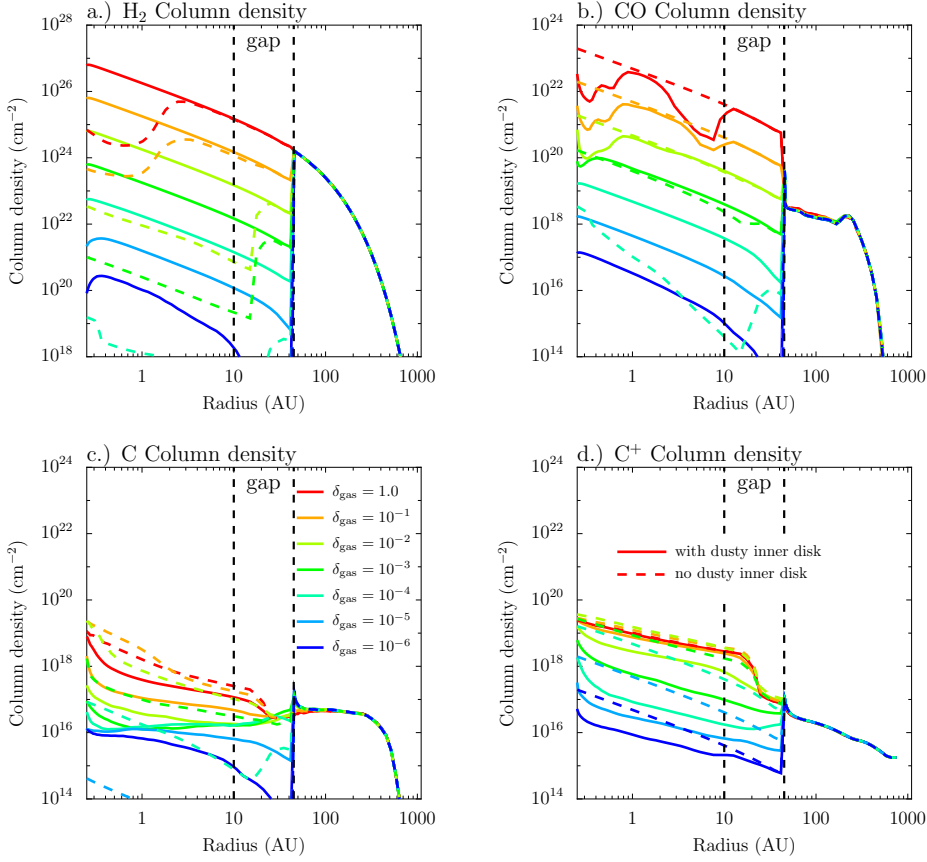}
\caption{Vertical column densities of the representative models (Table \ref{tab:param_tdisk}) \textbf{a)} H$_2$, \textbf{b)} CO, \textbf{c)} neutral carbon, and \textbf{d)} ionized carbon. Different colors shows models with different values of $\delta_{\rm gas}$. Solid lines represent models with a dusty inner disk present ($\delta_{\rm dust}=10^{-5}$) and the dashed lines models without dusty inner disk ($\delta_{\rm dust}=10^{-10}$).}
\label{fig:plot_column}
\end{figure*}

A first impression about the expected observables (images and intensities) can be obtained from column densities, since optically thin lines have intensities proportional to the column densities. In Figure \ref{fig:plot_column}, we present vertical column densities of H$_2$, CO, neutral carbon C and ionized carbon C$^+$. The figure gives both representative models (Table \ref{tab:param_tdisk}) with dusty inner disk ($\delta_{\rm dust}=10^{-5}$; solid lines) or without dusty inner disk ($\delta_{\rm dust}=10^{-10}$; dashed lines) and different amounts of gas in the cavity ($\delta_{\rm gas} = 10^{-6} - 1$). 

The H$_2$ column density of the model with dusty inner disk present (Figure \ref{fig:plot_column}a, solid line) follows the surface density profile for the model with $\delta_{\rm gas}=1$. In the cavity, the column density  scales linearly with $\delta_{\rm gas}$ down to $\delta_{\rm gas} \sim 10^{-4}$. Below this value, the signature of H$_2$ photodissociation can be seen by the column density decreasing quicker than the scaling with the amount of gas ($\delta_{\rm gas}$). The effect of photodissociation is first seen at the inner edge of the disk, because of strong irradiation and the dust free gap, and corresponds to the absence of H$_2$ formation on dust.  The models without dusty inner disk (Fig \ref{fig:plot_column}a, dashed line) show a different column density pattern and dependence on $\delta_{\rm gas}$. For $\delta_{\rm gas}=1$ and $10^{-1}$, not all hydrogen is in H$_2$ at radii $\lesssim 3$ AU (Sect. \ref{sec:innerdisk}). Outside this radius, the column density is very similar to the models with dusty inner disk present. For lower $\delta_{\rm gas}$, the amount of H$_2$ decreases quickly with $\delta_{\rm gas}$, because H$_2$ formation through H$^-$ gets inefficient. The H$_2$ column of the model with $\delta_{\rm gas}=10^{-3}$ in the inner 20 AU is two orders of magnitude lower compared to the model with dusty inner disk. Outside 20 AU, the column density is similar to the model with dusty inner disk due to the presence of PAHs. Differences in H$_2$ column density between models with and without dusty inner disk become larger for lower values of $\delta_{\rm gas}$. In the outer disk, the H$_2$ column density of all models is equal, independent of $\delta_{\rm gas}$ and the presence of a dusty inner disk.

The CO column density of the model with dusty inner disk (Figure \ref{fig:plot_column}b, solid line) does not follow the surface density profile. Two dips relative to the surface density profile (at $r < 1$ AU and $2 < r < 10$ AU) are due to carbon being bound in methane (Sect. \ref{sec:survival}). The drop in the CO column at 45 AU is for the same reason. The CO column density only follows the surface density profile outside 200 AU. The dips at $r<10$ AU are small in the model with $\delta_{\rm gas}=10^{-1}$ and the model with $\delta_{\rm gas}=10^{-2}$ has only the inner most dip ($r<1$ AU) present, as water does not freeze out anymore at $2 < r < 10$ AU. Models with lower $\delta_{\rm gas}$ have a more smooth column density profile, which approximately scales with $\delta_{\rm gas}$. As for H$_2$, only models with $\delta_{\rm gas}=10^{-5}$ and $10^{-6}$ show the signature of photodissociation. The column density pattern of the model without dusty inner disk is similar to that with inner disk for $\delta_{\rm gas} > 10^{-4}$, except that dips in the column density are missing. Below this value, CO is however dissociated more quickly compared to the model with dusty inner disk. As for H$_2$, we find again that the CO column in the outer disk is very similar for all models.

The C and C$^+$ columns are given in Figure \ref{fig:plot_column}c and \ref{fig:plot_column}d. Both species show a trend of decreasing column density with $\delta_{\rm gas}$. This is however much less pronounced than for CO and H$_2$. For example the C$^+$ column density of the models with dusty inner disk does only vary by about 3 orders of magnitude, while the column density of CO for the same series of models varies by more than 7 orders of magnitude. Approximately the same amount of variation in column density is found for C$^+$ (models without dusty inner disk) and C (models with inner disk). C shows larger variations in models without dusty inner disk , because for $\delta_{\rm gas} = 10^{-6}$, all carbon within the cavity can be ionized. The smaller variation in the C and C$^+$ column density compared to H$_2$ and CO results from the fact that these species exist in a transition layer on top of the midplane, which is found at different heights in the disk for different $\delta_{\rm gas}$, but still has similar column densities. The column densities of C and C$^+$ in the outer disk are comparable for all models. 

We conclude that the column densities of H$_2$ and CO within the cavity scale almost linearly with $\delta_{\rm gas}$ down to a value of $\delta_{\rm gas}$, where photodissociation becomes important. The column densities of C and C$^+$ on the other hand do not scale in this linear fashion.

\subsection{Line intensities and opacity effects} \label{sec:lineformation}

The CO rotational lines get optically thick above a certain column density and thus do not directly trace the gas mass in the gap or cavity anymore. An advantage of the CO rotational lines is that rare isotopologues ($^{13}$CO, C$^{18}$O, C$^{17}$O, $^{13}$C$^{18}$O, and $^{13}$C$^{17}$O) can be observed. The isotopologue lines have lower optical depth, due to their lower abundance and thus remain optically thin up to larger gas masses. In this section, we will study the integrated intensities and opacity effects on the example of CO $J=3-2$ and isotopologues. We discuss the representative model (Table \ref{tab:param_tdisk}) with dusty inner disk present ($\delta_{\rm dust}=10^{-5}$). Since $^{13}$C$^{17}$O is mostly below the detection limit and C$^{18}$O has a similar isotopologue ratio to C$^{17}$O, we mostly omit those two isotopologues in the following sections, but include them in the relevant figures.
  
\begin{figure*}[!htb]
\sidecaption
\includegraphics[width=0.66\hsize]{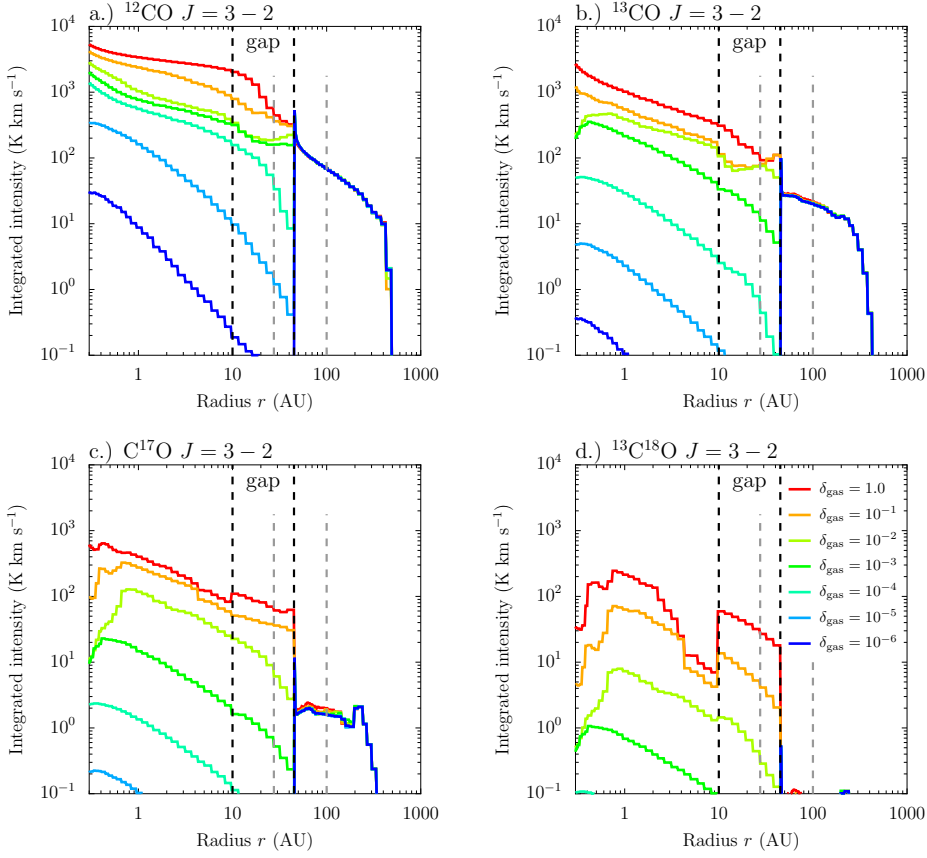}
\caption{Continuum subtracted integrated line intensity as function of the radius for CO $3-2$ and its isotopologues. The figure shows intensities derived from the representative models (Table \ref{tab:param_tdisk}). The vertical grey lines at 28 AU and 100 AU indicate the radii and which the intensities are provided in Figures \ref{fig:plot_fluxdeltagas} and \ref{fig:plot_fluxdeltagas2}.}
\label{fig:plot_flux_radius}
\end{figure*}

In Figure \ref{fig:plot_flux_radius}, the velocity integrated and continuum subtracted intensities depending on the radius are presented for a face-on disk. The gas mass in the cavity is varied with $\delta_{\rm gas}=10^{-6}-1$. The main isotopologue ($^{12}$CO) integrated intensity in the center of the gap (28 AU) decreases only by a factor of $\sim 4$ from $\delta_{\rm gas}=1$ to $\delta_{\rm gas}=10^{-3}$, due to the high optical depth of this line. For $\delta_{\rm gas} < 10^{-4}$, the line is optically thin and the integrated intensity scales almost linearly with the column density and thus $\delta_{\rm gas}$ (previous section). The lines of the rare isotopologues have lower optical depth and they show a linear dependence of integrated intensity with $\delta_{\rm gas}$ in the center of the gap (28 AU) already for $\delta_{\rm gas} \lesssim 10^{-1}$ (C$^{17}$O), $\delta_{\rm gas} \lesssim 10^{-2}$ ($^{13}$CO). $^{13}$C$^{18}$O does not get optically thick and scales linearly with $\delta_{\rm gas}$ in the gap. The rare isotopologue lines thus directly trace the column density and gas mass in the cavity already for much higher values of $\delta_{\rm gas}$ compared to $^{12}$CO.

\begin{figure}[!htb]
\center
\includegraphics[width=0.85\hsize]{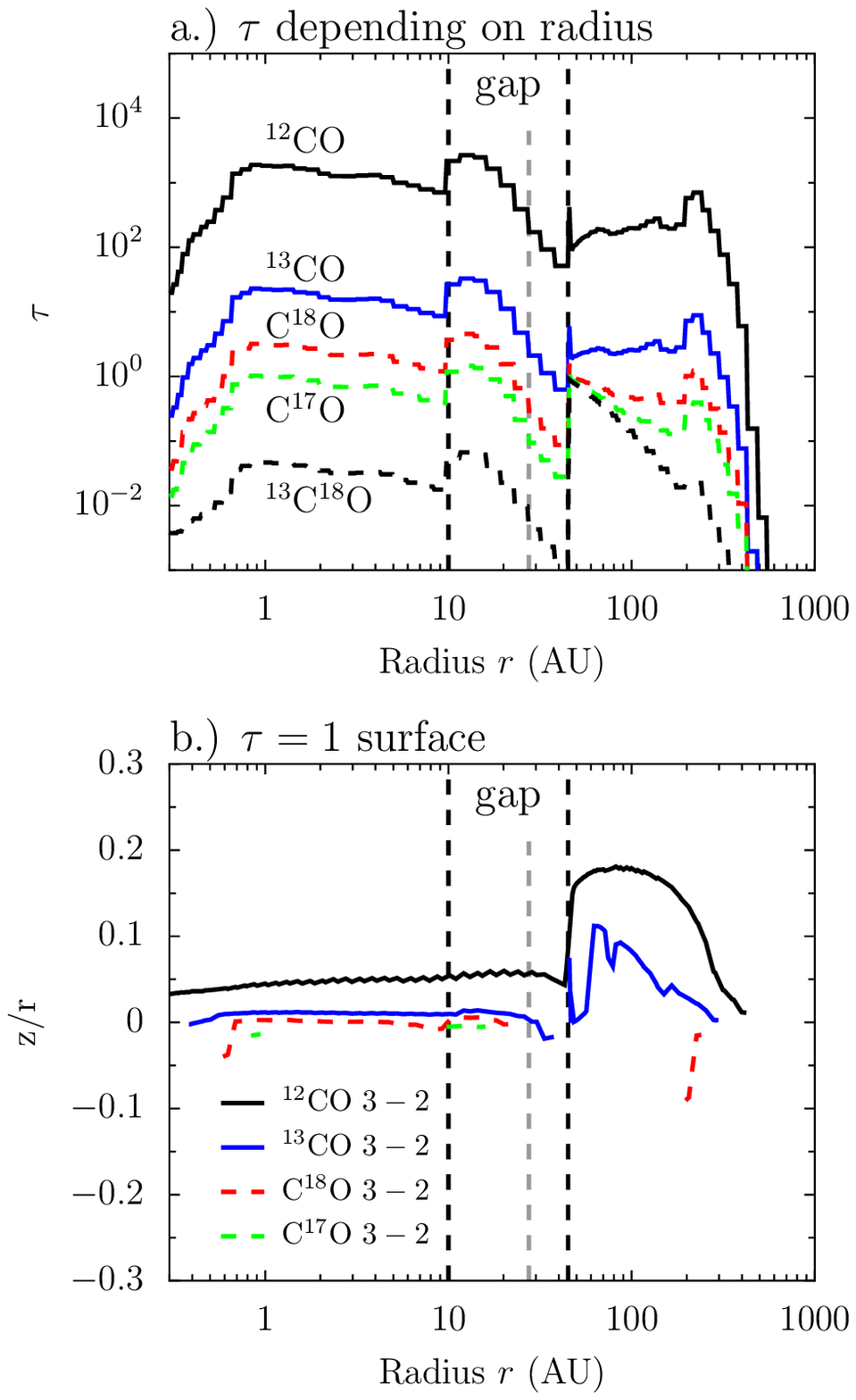}
\caption{Surface of $\tau=1$ and line opacities of CO $3-2$ and isotopologues for the representative model (Table \ref{tab:param_tdisk}) with $\delta_{\rm gas}=10^{-2}$ and a dusty inner disk present ($\delta_{\rm dust}=10^{-5}$). The vertical grey line at 28 AU shows the center of the gap. \textbf{Upper panel:} Line center opacity depending on the radius. \textbf{Lower panel:} Position of the $\tau=1$ surface for different radii.}
\label{fig:plot_tau}
\end{figure}

What is the optical depth of the lines? In Figure \ref{fig:plot_tau}a, the total line center opacity ($\tau = \tau_{\rm line} + \tau_{\rm dust}$) depending on the radius is shown for a face-on disk with $\delta_{\rm gas}=10^{-2}$. The line center opacities approximately scale with the column density and thus $\delta_{\rm gas}$. The dust opacity peaks in the outer disk, at the edge of the cavity ($R=45$ AU), with $\tau_{\rm dust} \sim 0.8$. Again looking at the center of the gap (28 AU), we find  $^{12}$CO line center opacities of a few 100. Using the scaling with $\delta_{\rm gas}$, we thus confirm that the line gets optically thin at $\delta_{\rm gas} \sim 10^{-4}$. Similarly, we see that C$^{17}$O reaches line center opacities of order 10 for $\delta_{\rm gas}=1$. $^{13}$C$^{18}$O remains optically thin even for $\delta_{\rm gas}=1$. 

The reason why optically thick lines still slightly increase with $\delta_{\rm gas}$ is, that the line can broaden or get optically thick higher in the disk, thus tracing warmer regions (\citealt{vanZadelhoff01}; \citealt{Dartois03}). What heights of the disk are traced by the CO lines? A simple way to answer this question is to look at the position, where $\tau=1$ is reached in vertical direction (BR12). In Figure \ref{fig:plot_tau}b the height of the $\tau=1$ surface is presented. The $\tau=1$ surface of $^{12}$CO is at considerably larger height compared to other isotopologues at all radii. Within the cavity, it is at approximately constant height, but it has a much larger height in the outer disk. The $^{13}$CO $\tau=1$ surface shows similar features however, deeper in the disk. The rare isotopologues C$^{18}$O and C$^{17}$O do not reach $\tau=1$ at all radii and have their $\tau=1$ surface close to the midplane or even on the far side of the disk. $^{13}$C$^{18}$O does not get optically thick. We conclude, that the main isotopologue line traces the warm upper atmosphere, while the rare isotopologues lines trace layers at lower heights in the disk or the whole column density, if they are optically thin.

In conclusion, line opacity effects are key for the understanding of the CO emission and their use as mass tracers. While the main isotopologue line is optically thick down to low gas mass in the cavity, the rare isotopologues can remain optically thin. Provided that their lines can be detected, combined observations of isotopologue lines thus allow us to directly trace a wide range of gas masses.

\subsection{Images} \label{sec:images}

\begin{figure*}[!htb]
\includegraphics[width=1.0\hsize]{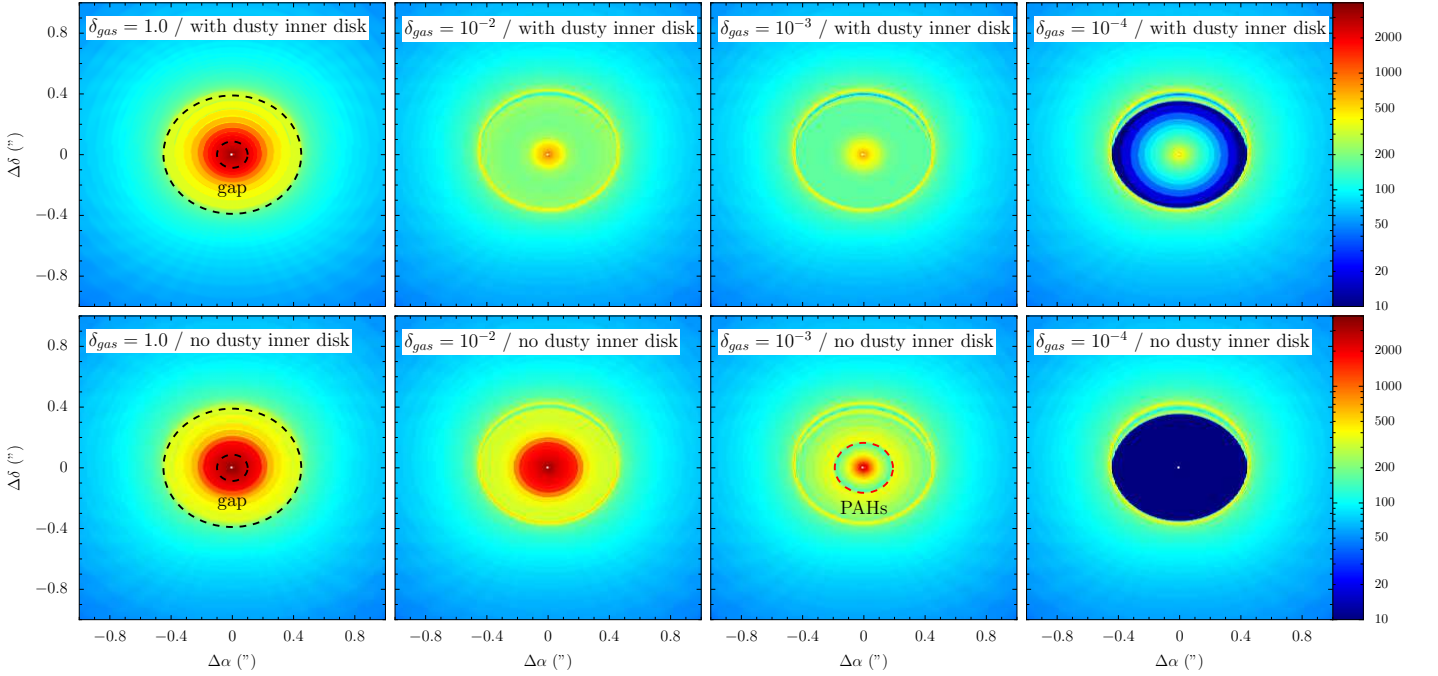}
\caption{Images of the integrated intensity (in K km s$^{-1}$) of $^{12}$CO $3-2$ from the series of representative models (Table \ref{tab:param_tdisk}). Models with $\delta_{\rm gas} = 1, 10^{-2}, 10^{-3}$ and $10^{-4}$ are shown. The inclination $i=30^\circ$ and the black dashed lines show the size of the gap ($10 < r< 45$ AU). Red dashed lines show the radius where PAHs can survive in absence of UV shielding by a dusty inner disk (Sect. \ref{sec:rep_physstructure}). \textbf{Upper panels:} Models with a dusty inner disk present ($\delta_{\rm dust}=10^{-5}$). \textbf{Lower panels:} Models without an inner disk ($\delta_{\rm dust}=10^{-10}$).}\label{fig:images_innerdisk}
\end{figure*}

\begin{figure*}[!htb]
\includegraphics[width=1.0\hsize]{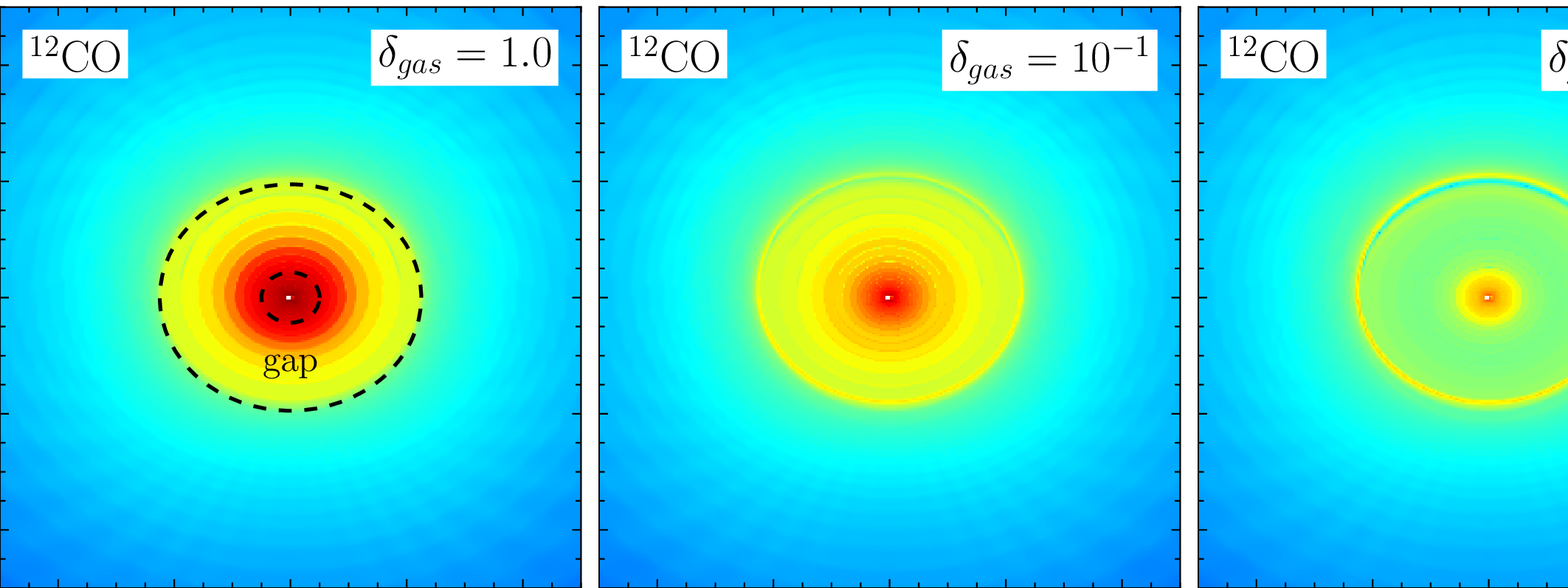}
\caption{Images of the integrated intensity (in K km s$^{-1}$) of CO isotopologues ($^{12}$CO, $^{13}$CO, C$^{17}$O, and $^{13}$C$^{18}$O) from the series of representative models (Table \ref{tab:param_tdisk}) with dusty inner disk present ($\delta_{\rm dust}=10^{-5}$). Models with $\delta_{\rm gas} = 1, 10^{-1}, 10^{-2}$ and $10^{-3}$ are shown. The inclination $i=30^\circ$ and the black dashed lines show the size of the gap ($10 < r< 45$ AU). The C$^{18}$O images look similar to the C$^{17}$O images and thus not shown. \textbf{1st row:} $^{12}$CO $3-2$. \textbf{2nd row:} $^{13}$CO $3-2$. \textbf{3rd row:} C$^{17}$O $3-2$. \textbf{4th row:} $^{13}$C$^{18}$O $3-2$.}\label{fig:plot_images_m14_inner}
\end{figure*}

\begin{figure*}[!htb]
\includegraphics[width=1.0\hsize]{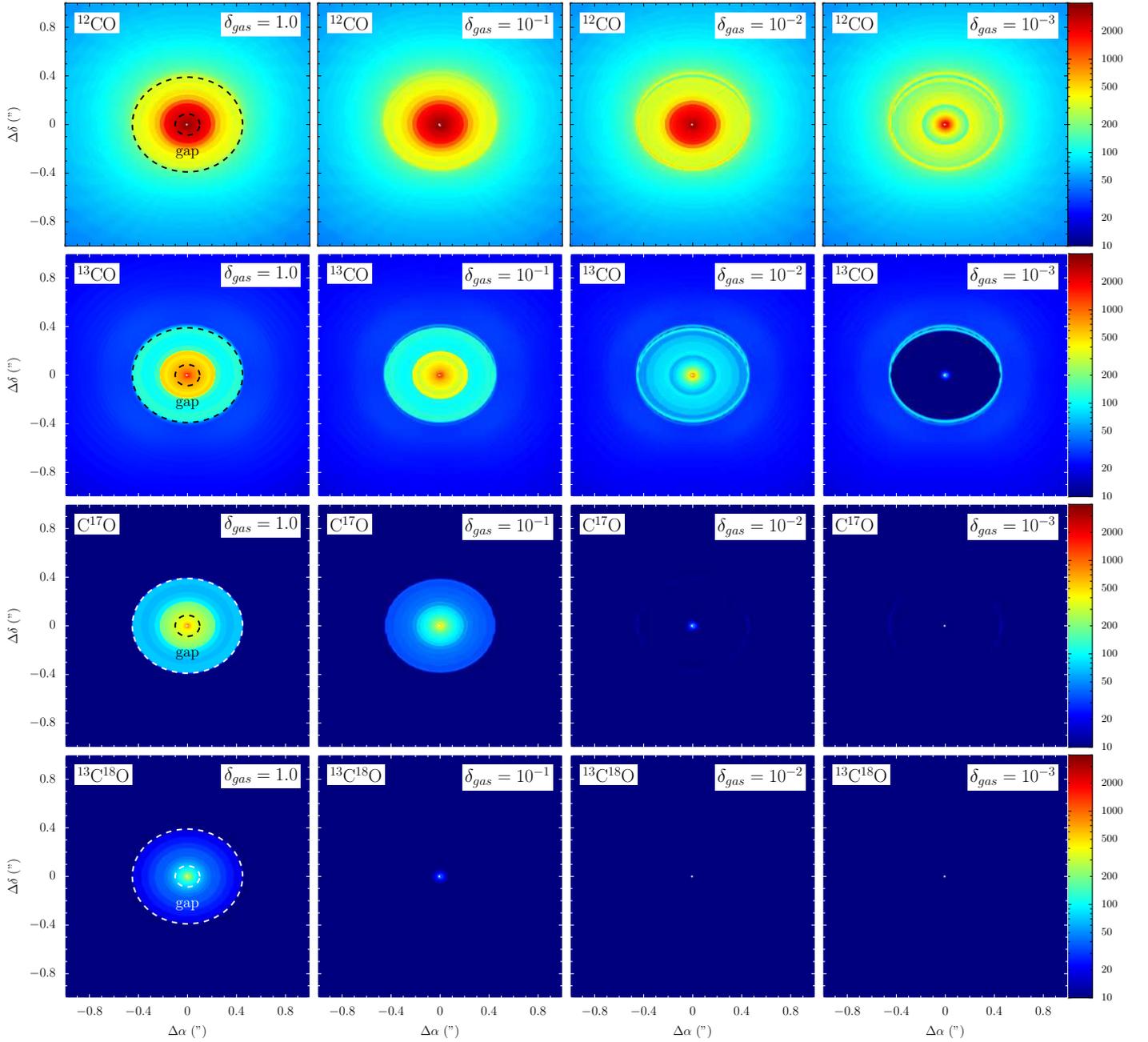}
\caption{Images of the integrated intensity (in K km s$^{-1}$) of CO isotopologues ($^{12}$CO, $^{13}$CO, C$^{17}$O, and $^{13}$C$^{18}$O) from the series of representative models (Table \ref{tab:param_tdisk}) without  dusty inner disk ($\delta_{\rm dust}=10^{-10}$). Models with $\delta_{\rm gas} = 1, 10^{-1}, 10^{-2}$ and $10^{-3}$ are shown. The inclination $i=30^\circ$ and the black dashed lines show the size of the gap ($10 < r< 45$ AU).  The C$^{18}$O images look similar to the C$^{17}$O images and thus not shown. \textbf{1st row:} $^{12}$CO $3-2$. \textbf{2nd row:} $^{13}$CO $3-2$. \textbf{3rd row:} C$^{17}$O $3-2$. \textbf{4th row:} $^{13}$C$^{18}$O $3-2$.}\label{fig:plot_images_m14_noinner}
\end{figure*}

As an example of line images derived from the models, Figure \ref{fig:images_innerdisk} presents (unconvolved) images of the ${}^{12}$CO $3-2$ integrated intensities derived from the series of models studied in the previous section. Compared are models with $\delta_{\rm gas}=1, 10^{-2}, 10^{-3}$ and $10^{-4}$ of both the series of models with or without dusty inner disk present. The distance is assumed to be 100 pc and the inclination is set to $30^\circ$. Intensities below 10 K km s$^{-1}$ are cut off, approximately corresponding to the detection limit of ALMA (see next section). The same figures for the $^{12}$CO $3-2$ line and isotopologues ($^{13}$CO, C$^{17}$O, and $^{13}$C$^{18}$O) comparing $\delta_{\rm gas}=10^{-3}-1$, either with or without dusty inner disk are given as Online-Figures (Figure \ref{fig:plot_images_m14_inner} and \ref{fig:plot_images_m14_noinner}).

The integrated intensities of CO and isotopologues in the cavity generally decreases with $\delta_{\rm gas}$. As discussed in the previous sections, this decrease is more pronounced towards lower $\delta_{\rm gas}$ ($< 10^{-3}$) which reflects the photodissociation of CO.

In some of the images, the warm upper edge of the outer disk wall can be seen. If the gas in the cavity is optically thick, only the edge pointed towards us can be seen (e.g. Figure \ref{fig:images_innerdisk}, images with $\delta_{\rm gas}=10^{-2}$). For an optically thin cavity also the edge of the far side is visible (e.g. Figure \ref{fig:images_innerdisk}, images with $\delta_{\rm gas}=10^{-4}$).

Models with or without a dusty inner disk look similar for $^{12}$CO and $\delta_{\rm gas}=1$. For lower $\delta_{\rm gas}$ and also the rare isotopologues, two mechanism acting in different directions are at work: Gas temperatures of models without inner disk are higher compared to those with inner disk (Sect. \ref{sec:innerdisk}), but in models without inner disk, the effect of photodissociation decreases the CO column density already for higher $\delta_{\rm gas}$ (Sect. \ref{sec:column}). Models without inner disk thus have higher integrated intensities in the cavity, as long as the CO column density is similar to models with inner disk. For the main isotopologue this is the case for $\delta_{\rm gas}$ down to $10^{-3}$ (Figure \ref{fig:images_innerdisk}). For $\delta_{\rm gas} = 10^{-4}$, however, the integrated intensity of the model without dusty inner disk is weaker, in agreement with the expectation from the CO column density (Figure \ref{fig:plot_column}). The rare isotopologues show the same effects, but their intensity starts dropping at higher $\delta_{\rm gas}$, due to lower optical depth. Differences between models with and without dusty inner disk increase to lower gas masses in the cavity.

%
%
\section{Discussion} \label{sec:discussions}

The gas mass in the gap of transition disks is a crucial parameter for theories of planet formation and measuring this parameter thus key. On the other hand, different mechanism for the formation of dust gaps would result in different amounts of gas in the gap. Thus, knowledge of the gas mass can also help to distinguish between different dust gap formation mechanism. The models presented in the previous section can be used to derive a variety of parameters, we will however focus on the use of CO lines, observed at high angular resolution by ALMA, as tracers of the gas mass in the cavity. We will also discuss, if the gas mass in the cavity can be derived from lower angular resolution observations or from atomic fine structure lines ([\ion{C}{I}], [\ion{C}{II}],[\ion{O}{I}]).

\subsection{Predictions for ALMA: CO} \label{sec:alma_co_pred}

In order to  detect low amounts of gas, a good sensitivity is crucial. In the following sections we will refer to the ALMA band 7 (CO $3-2$) detection limit as the line sensitivity for a $5\sigma$ detection in 1 hr on-source observation. Using the ALMA time sensitivity calculator\footnote{http://almascience.eso.org/call-for-proposals/sensitivity-calculator}, assuming typical weather conditions (3rd octile), dual polarization and the use of 50 12m antennas, the RMS noise in each 0.25 km s$^{-1}$ channel is 3 mJy (345 GHz), corresponding to a surface brightness of 3.1 K for a resolution of 0.1''. For a firm $5\sigma$ detection within 1 hr on-source observation and a beam small enough to resolve a typical cavity, an integrated intensity of order $\sim 7.7$ K km s$^{-1}$ is thus necessary (assuming an intrinsic line-width of order 1 km s$^{-1}$). 

\subsubsection{Tracing the gas mass in the cavity} \label{sec:tracegasmass}

\begin{figure}[htb]
\includegraphics[width=1.0\hsize]{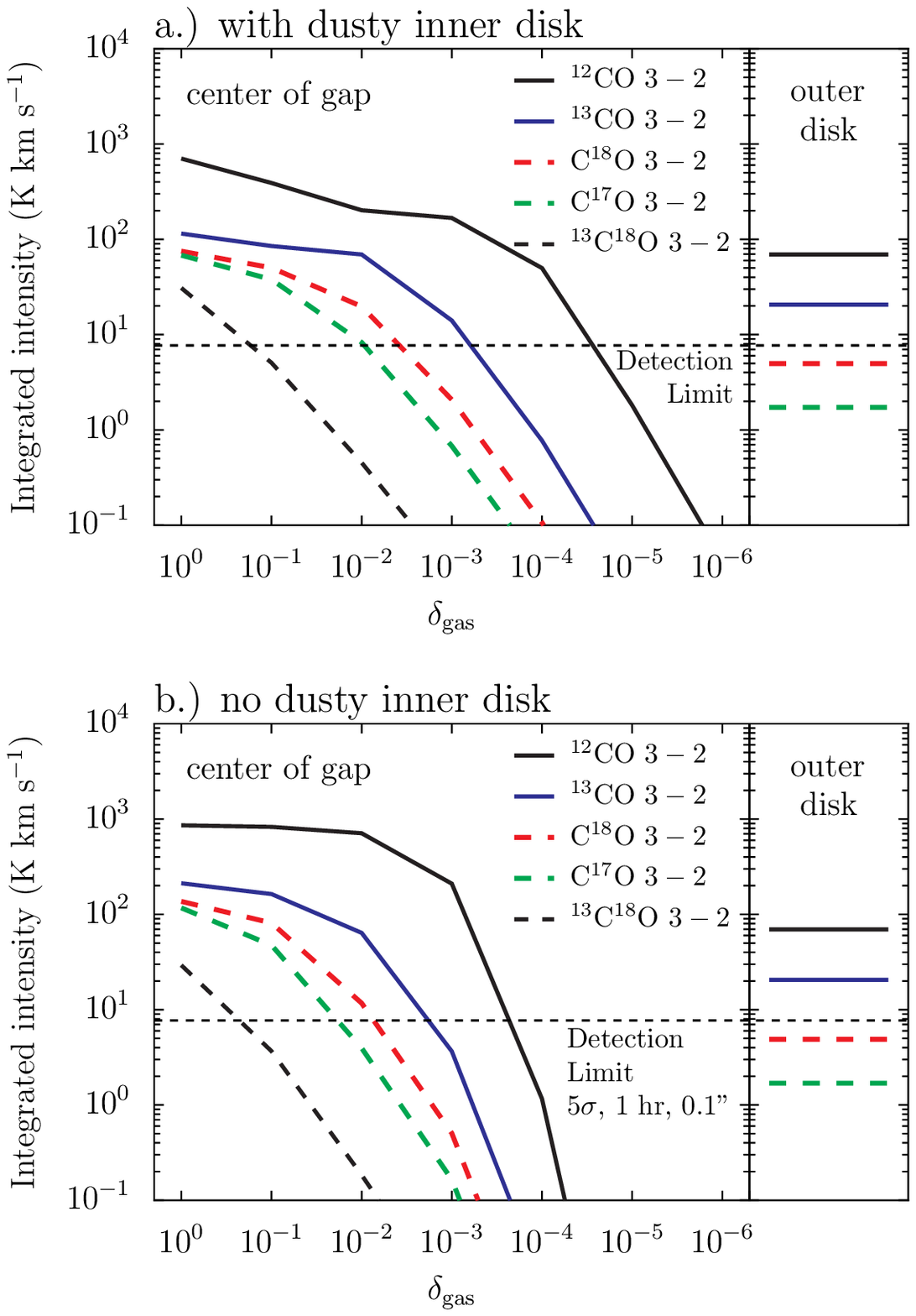}
\caption{Integrated intensities of CO $3-2$ and isotopologues at the center of the gap ($r=28$ AU) and in the outer disk ($r=100$ AU). The ALMA detection limit for a 5$\sigma$ detection after 1 hr on-source observation with a 0.1'' beam (see Section \ref{sec:discussions}) is given by the vertical dashed black line. \textbf{Upper panels:} Models with a dusty inner disk ($\delta_{\rm dust}=10^{-5}$). \textbf{Lower panels:} Models without inner disk ($\delta_{\rm dust}=10^{-10}$).}
\label{fig:plot_fluxdeltagas}
\end{figure}

What is the range of gas masses in the cavity that can be constrained with ALMA observations of CO rotational lines? Figure \ref{fig:plot_fluxdeltagas} shows the integrated intensities of $^{12}$CO $3-2$ and isotopologues in the center of the cavity (at $r=28$ AU) and the outer disk (at $r=100$ AU). Models with- and without dusty inner disk and for different values of $\delta_{\rm gas}$ are given. The integrated intensities at 100 AU of models with different $\delta_{\rm gas}$ are very similar ($\lesssim 20$ \%).

For a detection limit of order 7.7 K km s$^{-1}$, $^{12}$CO $3-2$ can be detected in the gap down to $\delta_{\rm gas} = 3 \times 10^{-5}$ in the case of a model with dusty inner disk. This corresponds to a gas mass in the gap of $3.6 \times 10^{-7}$ M$_\odot$ or 0.2 M$_{\rm Earth}$ (Table \ref{tab:param_mass}). For models without dusty inner disk, the detection limit in gas mass is about an order of magnitude higher ($\delta_{\rm gas}=3 \times 10^{-4}$) and corresponds to 2 M$_{\rm Earth}$. For both series of models with and without dusty inner disk, the $^{12}$CO integrated intensity in the gap stays within a factor of a few for high values of $\delta_{\rm gas}$, where the line is optically thick. The main isotopologue $^{12}$CO line is optically thick for $\delta_{\rm gas} > 10^{-4}$ (with dusty inner disk) or $\delta_{\rm gas} > 10^{-3}$ (no dusty inner disk). The rare isotopologues get only optically thick for higher values of $\delta_{\rm gas}$, with for example $^{13}$CO being optically thick for $\delta_{\rm gas} \gtrsim 10^{-2}$ (Sect. \ref{sec:lineformation}). Above $\delta_{\rm gas}=10^{-1}$, even C$^{17}$O gets optically thick, but in this mass range, $^{13}$C$^{18}$O is optically thin and above the detection limit. For a disk with $\delta_{\rm gas}=1$, $^{13}$C$^{18}$O can be detected on a $20 \sigma$ level after 1 hr on-source integration.

We conclude that combined observations of different isotopologues are key and will allow us to directly trace a wide range of masses from a full disk with $\delta_{\rm gas}= 1$ down to the detection limit of the main isotopologue at values of $\delta_{\rm gas} \sim (0.3 - 3) \times 10^{-4}$ which are $3.5-4.5$ orders of magnitude lower in mass. Differences between models with or without dusty inner disk are smaller for higher gas masses, because the lines are optically thick in the cavity and increase to lower masses, where the lines get optically thin.

The same analysis for $^{12}$CO $6-5$ (691 GHz) and isotopologues shows very similar results and is not reported here. The flux sensitivity of ALMA band 9 for the same conditions\footnote{1st octile weather would reduce the values to 19 mJy and thus 12 K km s$^{-1}$} is with 40 mJy a factor of 13 higher, leading to a $5\sigma$ detection limit in a 0.1'' beam after an 1 hr on-source of 26 K km s$^{-1}$. Since the CO $6-5$ integrated intensities are similar ($< 50$ \% difference in brightness temperature compared to CO $3-2$, the CO $3-2$ line is more suitable to trace lower amounts of gas. For transition disks with a small cavity radius, the factor of 2 smaller beam of CO $6-5$ may be required, though.

The gas temperature in the cavity is a few 100 K (Sect. \ref{sec:rep_physstructure}) which is high enough to excite the first vibrational levels of CO at $E_u \gtrsim 3100$ K to some extent. However, due to the combination of high critical density ($\sim 10^{14}$ cm$^{-3}$) and high upper level energy of the rovibrational transitions, they are more sensitive to excitation conditions (density and gas temperature) compared to pure rotational lines. In particular the IR lines depend very much on the gas temperature which is known to be uncertain in thermo-chemical models. In addition, if some dust still remains in the gap, the near infrared lines are more prone to dust extinction than the submillimeter lines. We conclude that submillimeter lines are more suited to trace the gas mass in the cavity.

\subsubsection{The need for high resolution interferometry} \label{sec:needhighresol}

Since the gas temperature and column densities generally increase to smaller radii, the intensity increases to smaller radii (Figure \ref{fig:plot_flux_radius}). This raises the question, if low-resolution (single-dish) observations are sufficient to the trace the gas in the inner disk, due to the potentially higher intensity from gas inside the cavity? 

\begin{figure*}[!htb]
\sidecaption
\includegraphics[width=0.75\hsize]{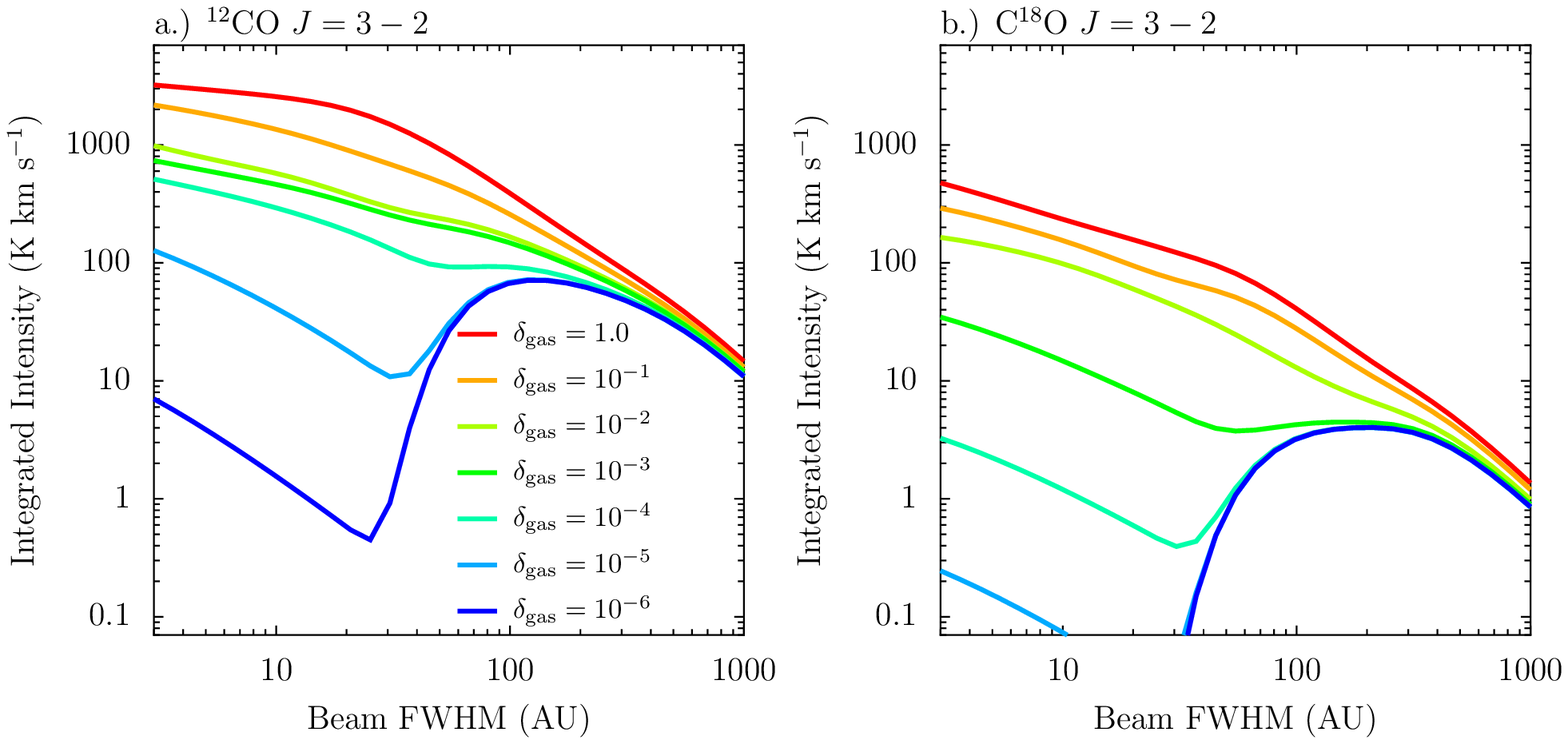}
\caption{Beam convolved integrated intensities of CO $3-2$ for different beam sizes. The representative model with a dusty inner disk present is shown. The disk is face on. \textbf{Left panel:} $^{12}$CO $3-2$ \textbf{Right panel:} C$^{18}$O $3-2$.}
\label{fig:images_beam_conv}
\end{figure*}

To answer this question, Figure \ref{fig:images_beam_conv} shows the beam convolved integrated intensities of $^{12}$CO $3-2$ and C$^{18}$O $3-2$ for different beam sizes. The disk is face on and has a dusty inner disk present. Other isotopologues show the same features and are thus not given. Beam sizes with FWHM larger than a few 100 AU (a few arcseconds at 100 pc) thus result in integrated intensities within a factor of a few, because the emission of the optically thick lines is quickly dominated by the much larger area from the outer disk and optically thin lines by the larger mass from the outer disk. Single-dish observations are thus clearly insufficient to draw constraints on the gas mass in the cavity.

\begin{figure}[!htb]
\includegraphics[width=1.0\hsize]{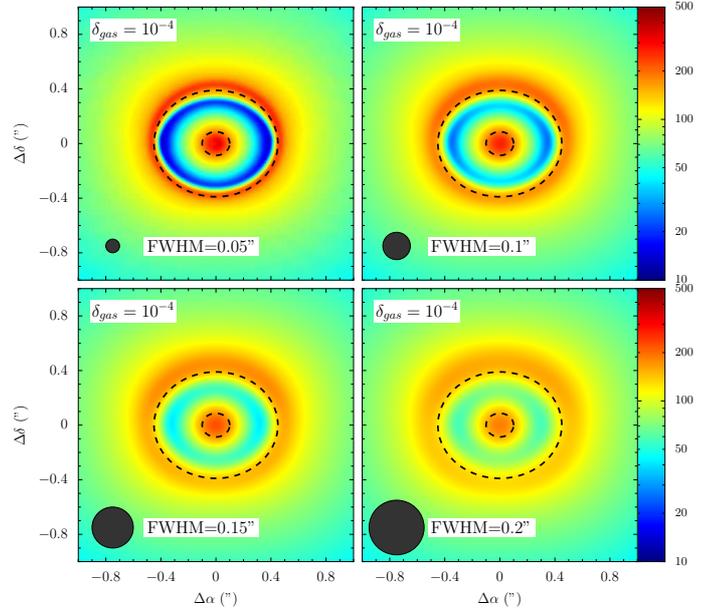}
\caption{Beam convolved images of the integrated intensity (in K km s$^{-1}$) of $^{12}$CO $3-2$ from a representative model (Table \ref{tab:param_tdisk}) with dusty inner disk present and $\delta_{\rm gas}=10^{-4}$ (Sect. \ref{sec:survival}). The inclination is $i=30^\circ$ and the images are convolved to a Gaussian beam with FWHM 0.05'', 0.1'', 0.15'' and 0.2''. The black dashed lines indicate the size of the gap ($r=10-45$ AU).}
\label{fig:images_beam}
\end{figure}

What resolution is necessary to constrain the gas mass in the cavity? Figure \ref{fig:images_beam} shows images of the model with dusty inner disk present ($\delta_{\rm dust}=10^{-5}$) and $\delta_{\rm gas}=10^{-4}$ convolved to different beam sizes (FWHM 0.05'', 0.1'', 0.15'' and 0.2''). The minimum integrated intensity inside the gap differs between the beams. Only in the case of the two smallest beams (0.05'' and 0.1''), the integrated intensity drops to a value $< 10$ K km s$^{-1}$ found in the unconvolved image (Figure \ref{fig:images_innerdisk}). Due to the strong intensity gradients of gap to inner disk and gap to outer disk, a small beam size is crucial to derive correct gas masses. For models without strong gradients at the edge of the gap, this issue is less severe (e.g. $^{12}$CO $3-2$ with dusty inner disk and $\delta_{\rm gas} > 10^{-4}$). We conclude that a beam size of a fraction of the gap is crucial to derive the gas mass.

\subsubsection{Dependence on parameters} \label{sec:paramdep}

\begin{figure*}[htb]
\includegraphics[width=0.95\hsize]{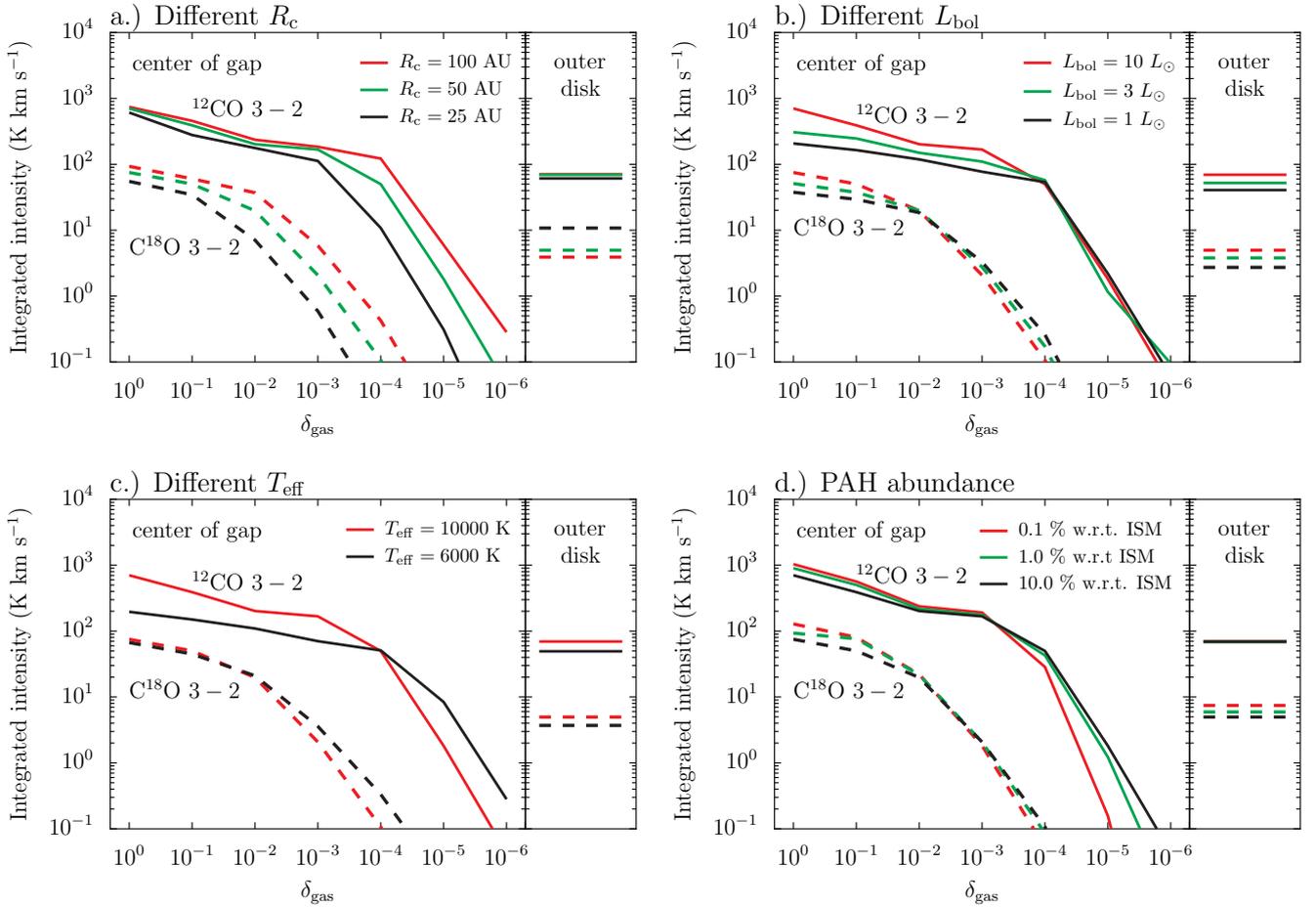}
\caption{Integrated intensities of $^{12}$CO $3-2$ (solid lines) and C$^{18}$O $3-2$ (dashed lines) at the center of the gap ($r=28$ AU) and in the outer disk ($r=100$ AU). The panels show the intensities for variations of one parameter ($R_{\rm c}$, $L_{\rm bol}$, $T_{\rm eff}$ and the PAH abundance). If not varied, the parameters are set to the values of the representative models (Table \ref{tab:param_tdisk}: $R_{\rm c}=50$ AU, $L_{\rm bol}=10$ $L_\odot$, $T_{\rm eff}=10000$ K and a PAH abundance of 10 \% the ISM abundance).}
\label{fig:plot_fluxdeltagas2}
\end{figure*}

How do the CO integrated intensities in the gap depend on the the other parameters of the grid of models, such as the disk mass (parameterized by $R_{\rm c}$), the bolometric luminosity ($L_{\rm bol}$), the stellar spectrum ($T_{\rm eff}$), and the PAH abundance? In Figure \ref{fig:plot_fluxdeltagas2} the $^{12}$CO $3-2$ and C$^{18}$O $3-2$ integrated intensities at the center of the gap and the outer disk are shown. One of the parameter ($R_{\rm c}, L_{\rm bol}, T_{\rm eff}$, or the PAH abundance) is varied, while the others are set to the value of the  representative model (Table \ref{tab:param_tdisk}). The dependence of the other isotopologues is similar and thus not shown. Since the X-ray spectrum is not found to be an important parameter for the CO integrated intensity in the gap, it is discussed in Appendix \ref{sec:app_xray}.

Increasing $R_{\rm c}$ from 50 AU to 100 AU leads to a factor of 2.5 larger gas mass in the cavity  (gap and inner disk) for a given $\delta_{\rm gas}$ (Table \ref{tab:param_mass}). Conversely, $R_{\rm c}=25$ AU yields a factor of 2.8 lower gas mass compared to $R_{\rm c}=50$. For $\delta_{\rm gas} \gtrsim 10^{-3}$, the integrated intensity of $^{12}$CO in the gap agrees to within a factor of $<2$ between models with different $R_{\rm c}$. This results from line opacity effects in combination with similar temperatures at the height in the disk, which is probed by this line. Below $\delta_{\rm gas}=10^{-3}$, the integrated intensity of $^{12}$CO from models with lower disk mass drop in intensity at higher values of $\delta_{\rm gas}$, compared to those with lower disk mass. The drop scales about linear with the disk mass and thus similar values of the gas mass in the cavity can be detected. This reflects that the density structure in the cavity in more massive models with a lower value of $\delta_{\rm gas}$ is similar to models with less mass, but higher $\delta_{\rm gas}$ (Figure \ref{fig:allsurfs}). The C$^{18}$O integrated intensity shows similar features as $^{12}$CO. However, the C$^{18}$O integrated intensity of models with different $R_{\rm c}$ are only similar for $\delta_{\rm gas} \gtrsim 10^{-2}$, where C$^{18}$O is optically thick.

Higher bolometric luminosities lead to higher gas and dust temperatures. This is seen in the increased $^{12}$CO integrated intensity for higher $L_{\rm bol}$ and $\delta_{\rm gas} > 10^{-4}$. In this mass range, the $^{12}$CO line is optically thick. Likewise, C$^{18}$O shows the same dependence for $\delta_{\rm gas} > 10^{-2}$. Below $\delta_{\rm gas}=10^{-2}$, the C$^{18}$O lines become optically thin and trace the column density. Due to increased photodissociation for higher $L_{\rm bol}$, the CO column density generally decreases with $L_{\rm bol}$. Below $\delta_{\rm gas}= 10^{-4}$, the dependence of CO column density with $L_{\rm bol}$ is more complex, as the integrated intensity of $^{12}$CO shows. In this range of $\delta_{\rm gas}$, additional CO formation due to higher gas temperatures can affect the total column density.

A harder spectrum of the star (higher $T_{\rm eff}$) has different effects on the chemistry and heating/cooling. This is discussed in detail in BR12 in the context of high-$J$ CO lines. Here, a harder spectrum with more CO/H$_2$ dissociating photons and also more heating (e.g. through the photoelectric effect) leads to higher $^{12}$CO intensities in the gap, if the line is optically thick ($\delta_{\rm gas} > 10^{-4}$). For lower values of $\delta_{\rm gas}$, a harder spectrum leads to more photodissociation and thus lower $^{12}$CO intensities. Since the C$^{18}$O line forms deeper in the atmosphere (Sect. \ref{sec:lineformation}), the effect of the increased gas temperature in the upper atmosphere is not seen. When C$^{18}$O is optically thin ($\delta_{\rm gas} < 10^{-2}$), the decreased integrated intensity for higher $T_{\rm eff}$ is also a result of the increased photodissociation.

PAHs affect the CO integrated intensity in the gap through a combination of chemical effects (e.g. formation of H$_2$ on their surface), heating by the photoelectric effect and absorption of FUV radiation. Lower PAH abundances lead to less H$_2$ formation and heating, but also less FUV absorption. Thus, CO forms deeper in the disk. Since CO is an important coolant in the gap (Sect. \ref{sec:gastemp}), the gas temperature at the top layer where CO forms is slightly higher. For high values of $\delta_{\rm gas}$, where the $^{12}$CO or C$^{18}$O are optically thick and trace this top layer, lower PAH abundances thus lead to higher integrated intensities in the gap. On the other hand, for low values of $\delta_{\rm gas}$, where the lines are optically thin, the additional photodissociation due to less FUV shielding by PAHs and less efficient H$_2$ formation leads to lower amounts of CO and thus lower integrated intensities.

We conclude that the main parameter for the low-$J$ CO integrated line intensity is the amount of gas within the cavity ($\delta_{\rm gas}$). Other parameters change the gas mass at which CO drops below the detection limit by only a factor of up to a few.

\subsection{Predictions for ALMA: C} \label{sec:carbonI}

\begin{figure}[htb]
\includegraphics[width=1.0\hsize]{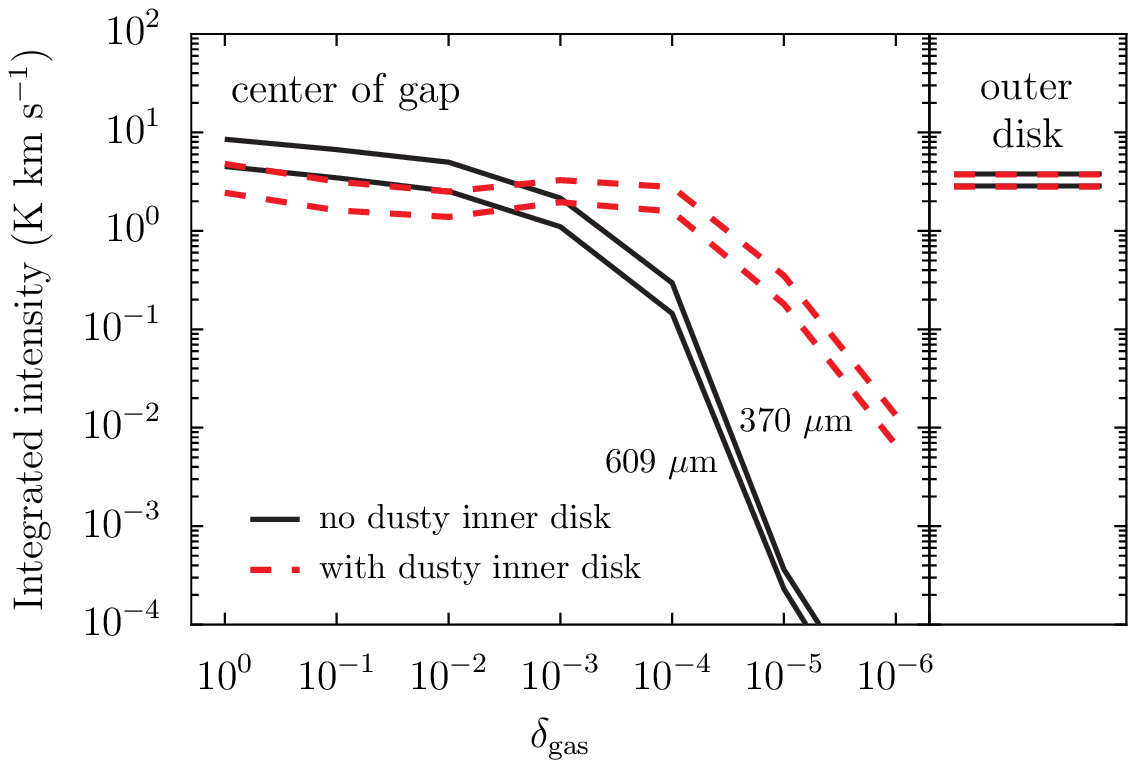}
\caption{Integrated intensities of [\ion{C}{I}] $^3$P$_2$-$^3$P$_1$ at 370 $\mu$m and $^3$P$_1$-$^3$P$_0$ at 609 $\mu$m at the center of the gap ($r=28$ AU) and in the outer disk ($r=100$ AU). Models with a dusty inner disk ($\delta_{\rm dust}=10^{-5}$) are shown in solid line, models without inner disk ($\delta_{\rm dust}=10^{-10}$) in dashed line. The 370 $\mu$m line has higher integrated intensities than the 609 $\mu$m line for the same $\delta_{\rm gas}$.}
\label{fig:plot_fluxdeltagas_cI}
\end{figure}

Neutral carbon lines are other potential tracers of gas in cavities (Sect. \ref{sec:atomcarbon}). However, neutral carbon is in a layer between the ionized carbon in the upper atmosphere and molecular gas (CO) closer to the midplane and has similar column densities for different $\delta_{\rm gas}$ (Figure \ref{fig:plot_column}c). Thus, the [\ion{C}{I}] lines observable by ALMA are not expected to be very sensitive on the gas mass. Figure \ref{fig:plot_fluxdeltagas_cI} shows the intensity in the gap and the outer disk depending on $\delta_{\rm gas}$ for the representative models (Table \ref{tab:param_tdisk}) with or without dusty inner disk.

The two carbon lines indeed do not change considerably with $\delta_{\rm gas}$ for $\delta_{\rm gas} \geq 10^{-4}$ (models with dusty inner disk) and $\delta_{\rm gas} \geq 10^{-3}$ (models without dusty inner disk). The intensities are much lower compared to CO and at higher frequency, which are more difficult to observe. The ALMA time sensitivity estimator for the same settings as above, shows that for the same detection limit, about a factor of 10 longer integration time is needed. Even this sensitivity is only marginal sufficient to detected [\ion{C}{I}] at the necessary angular resolution. We conclude that [\ion{C}{I}] observation are less suited to measure the amount of gas in cavities of transition disks.

\subsection{Tracing transition disks with Herschel: C$^+$?} \label{sec:herschel}

The fine structure line of ionized carbon ([\ion{C}{II}] $^2$P$_{3/2}$-$^2$P$_{1/2}$ at 158 $\mu$m) cannot be studied from the ground. The Herschel Space Observatory (\citealt{Pilbratt10}) has observed the line towards different transition disks with the PACS instrument (\citealt{Poglitsch10}) and velocity resolved with the HIFI instrument (\citealt{deGraauw10}). Due to the better sensitivity of PACS compared to HIFI, most observations have been carried out spectrally unresolved with PACS and we focus here on integrated intensities. The angular resolution of Herschel at 158 $\mu$m is $\sim 11''$, corresponding to $\sim 1000$ AU at 100 pc. Cavities can thus only be detected indirectly, if the emission from the cavity is much stronger compared to the outer disk and dominates the unresolved flux.

\begin{figure}[!htb]
\center
\includegraphics[width=0.8\hsize]{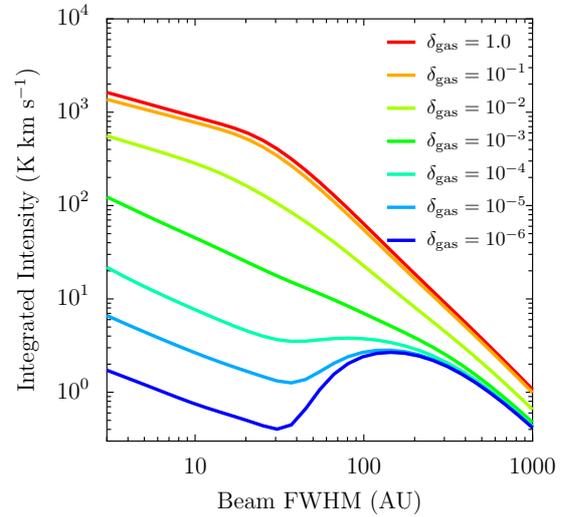}
\caption{Beam convolved integrated intensities of [\ion{C}{II}] $^2$P$_{3/2}$-$^2$P$_{1/2}$ at 158 $\mu$m for different beam sizes. The disk is face on and has a dusty inner disk present.}
\label{fig:images_beam_conv_cii}
\end{figure}

Figure \ref{fig:images_beam_conv_cii} shows the integrated intensity of [\ion{C}{II}] for different beam sizes. The representative models (Table \ref{tab:param_tdisk}) including a dusty inner disk present are shown. Similar to the equivalent figure for CO (Figure \ref{fig:images_beam_conv}), a large increase in integrated intensity with $\delta_{\rm gas}$ is found for small beam sizes compared to the size of the cavity. If the beam size is much larger and a substantial fraction of the intensity is due to the outer disk, the integrated intensity depends little on $\delta_{\rm gas}$ (less than a factor of 3). The increase is larger for models with $\delta_{\rm gas}=10^{-1}$ and $1$, but models with lower values of $\delta_{\rm gas}$ agree to within $\sim 50$ \% at beam sizes much larger than the cavity. Good knowledge about the outer radius of the disk would be required to draw conclusions at this level and we conclude that only for high amounts of gas in the cavity, some of the spatially unresolved [\ion{C}{II}] emission can be due to emission from within the cavity.

\begin{figure*}[!htb]
\center
\includegraphics[width=0.95\hsize]{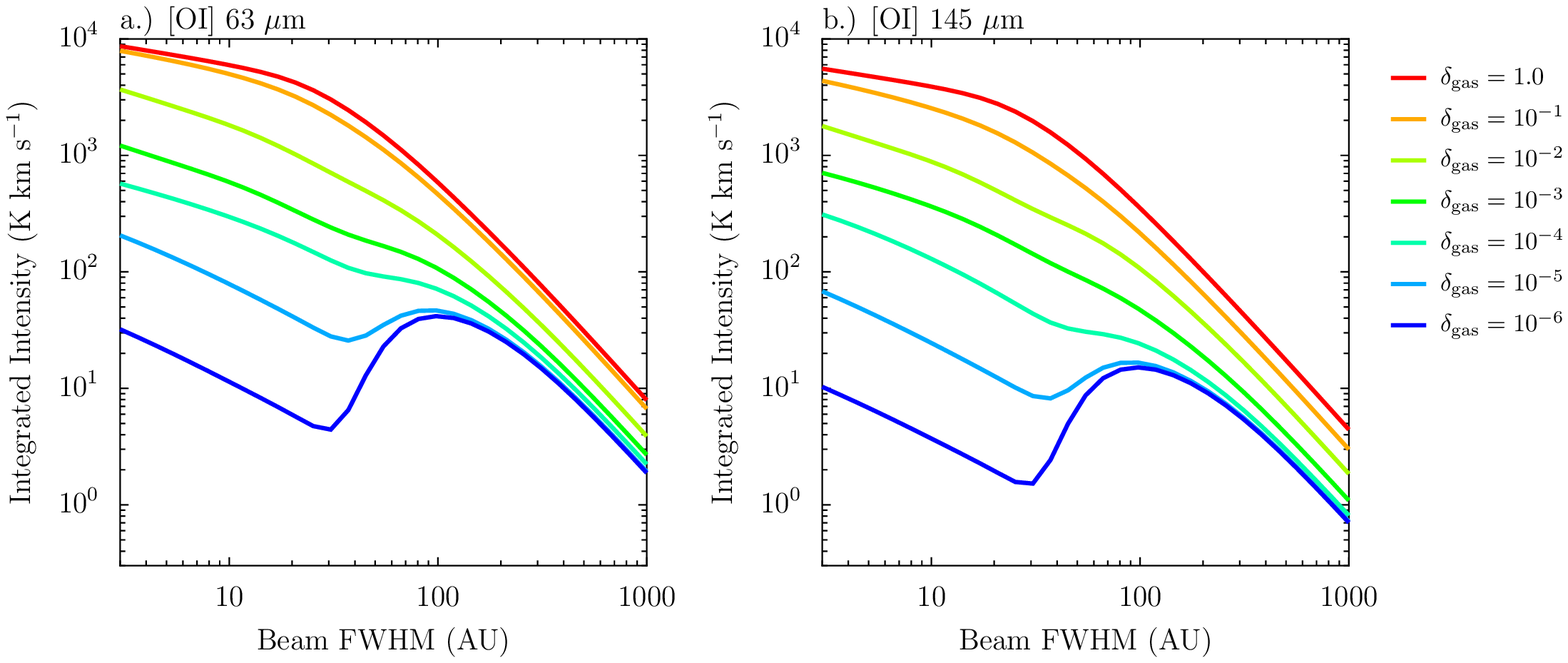}
\caption{Beam convolved integrated intensities of [\ion{O}{I}] $^3$P$_1$ - $^3$P$_2$ at 63 $\mu$m and  $^3$P$_0$ - $^3$P$_1$ at 145 $\mu$m and for different beam sizes. The disk is face on and has a dusty inner disk present.}
\label{fig:images_beam_conv_oi}
\end{figure*}

For completeness, the two neutral oxygen fine structure lines at 63 $\mu$m and 145 $\mu$m observed by the PACS instrument onboard Herschel are also discussed. Compared to the lines of CO, [\ion{C}{I}] and [\ion{C}{II}] studied before, the [\ion{O}{I}] lines have a higher upper level energy of 228 K and 327 K, respectively (compared to $< 100$ K). The oxygen emission is thus more centrally peaked (see BR12) and the cavity contribution to spatially unresolved observations should be larger compared to [\ion{C}{II}]. Figure \ref{fig:images_beam_conv_oi} (online only) shows the [\ion{O}{I}] integrated intensities in different beams. For beams corresponding to FHWM larger than a few 100 AU, the increase in intensity for the models with $\delta_{\rm gas} = 10^{-1}$ and $1$ is indeed larger compared with [\ion{C}{II}] (up to about a factor 6). Models with lower $\delta_{\rm gas}$ however still agree to within a factor of 2-3 and measuring gas masses in the cavity using [\ion{O}{I}] lines is challenging as well. 

\subsection{Implications} \label{sec:implications}

The previous sections have shown that combined observations of CO rotational lines and isotopologues allow us to directly trace a wide range of gas masses in the cavity corresponding to a full gas disk ($\delta_{\rm gas}=1$) down to $\delta_{\rm gas} \sim (0.3 - 3) \times 10^{-4}$. Even lower values of $\delta_{\rm gas}$ may be traced with observations longer than 1 hr on-source.

How can this range of gas masses help to distinguish between different dust gap formation mechanism? Clearing of the gap by a single Jupiter mass planet reduces the gas mass in the gap by a factor of 10-1000 ($\delta_{\rm gas}=10^{-1} - 10^{-3}$), depending on the viscosity. A more massive 9 M$_{\rm Jupiter}$ planet leads to $\delta_{\rm gas} \sim 10^{-2} - 10^{-3}$ (\citealt{Pinilla12b}). Photoevaporation (e.g. \citealt{Bally82,Hollenbach94,Clarke01,Alexander06a,Ercolano08,Gorti09}) has short processing time-scales and once its effect sets in, would quickly lead to a gas free cavity. Pure grain-growth on the other hand  (e.g. \citealt{Dullemond05,Birnstiel12b}) should not alter the gas structure much ($\delta_{\rm gas} \sim 1$).

Thus, high gas masses ($\delta_{\rm gas} \gtrsim 10^{-1}$) point to gap clearing by a less massive planet ($\lesssim$ 1 M$_{\rm Jupiter}$) or pure grain-growth. Resolved submillimeter dust images may help to distinguish between the two scenarios, because clearing by a planet results in a gap, while grain growth leads to a hole empty of submm grains. We note however, that models of pure grain growth are unable to explain the gap formation alone (\citealt{Birnstiel12b}). Intermediate values of $\delta_{\rm gas}$, with $\delta_{\rm gas} < 10^{-1}$, point to planet-disk interaction. In this mass regime, knowledge of the gas mass in the gap is of special interest, as it sets constraints for further planet formation or the dynamics of the planets. To estimate the planetary mass, additional constrains on the viscosity may help to break the degeneracy between reduction of the gas mass in the gap and viscosity (\citealt{Mulders13}). Low gas masses in the gap ($\delta_{\rm gas} < 3 \times 10^{-5}$) on the other hand are hard to achieve with planet clearing, even in the case of multiple systems (\citealt{DodsonRobinson11,Zhu11}). Such low masses thus point to photoevaporation, possibly in combination with planet clearing (\citealt{Rosotti13}). We conclude that measuring the gas mass in the cavity alone is not sufficient to distinguish between different gap formation mechanism or to constrain details of the mechanism at work. It can however exclude some mechanisms and help to confirm or discard constraints by other tracers (e.g. continuum images or the SED). In a forthcoming paper, we plan to couple the present model with a model for dust evolution (\citealt{Birnstiel10,Birnstiel11,Birnstiel12}) which will allow us to study the combined use of lines and the continuum in order to better constrain the gap formation mechanism.

\section{Conclusions and outlook} \label{sec:conclusions}

In this work, we have applied a new type of thermo-chemical models on transition disks to study the survival of simple molecules (H$_2$ and CO) in the dust free gap of these disks for a range of parameters. Predictions for ALMA are derived and observation strategies to trace the amount of gas in the cavity are discussed. While the focus of the work is on high-resolution observations that can trace the cavity spatially resolved, we also discuss the possibility to trace the gas mass in the cavity using spatially unresolved Herschel observations.\\

Our main conclusions are

\begin{enumerate}
\item CO and H$_2$ can survive the intense stellar UV radiation in gaps of transition disks down to gas masses in the gap corresponding to a fraction of the Earth mass ($\sim 0.004$ M$_{\rm Earth}$). Important for the survival of CO and H$_2$ is self- and mutual-shielding and alternative H$_2$ formation routes. Alternative H$_2$ formation routes are formation on PAHs or through H$^-$, which can both proceed in gas depleted of dust (Sect. \ref{sec:survival}).
\item The main parameter for the CO emission from the cavity is the gas mass in the cavity. The integrated intensity of optically thin CO isotopologue lines scales linearly with gas mass down to the mass where photodissociation gets important. Other parameters such as the outer disk mass, bolometric luminosity, shape of the stellar spectrum or PAH abundance are less important (Sect. \ref{sec:column}, \ref{sec:tracegasmass}, \ref{sec:paramdep}).
\item The lines of the main isotopologue $^{12}$CO are optically thick for $\delta_{\rm gas} > 10^{-4}$, where $\delta_{\rm gas}$ is the scaling of the gas mass in the cavity relative to a full gas disk. The rare isotopologues are already optically thin for $\delta_{\rm gas} \lesssim 10^{-1}$ (C$^{17}$O), $\delta_{\rm gas} \lesssim 10^{-2}$ (C$^{18}$O and $^{13}$CO). $^{13}$C$^{18}$O remains optically thin for $\delta_{\rm gas}=1$, and is still detectable for high gas masses (Sect. \ref{sec:lineformation}, \ref{sec:tracegasmass}).
\item Combined observations of CO isotopologues ($^{12}$CO, $^{13}$CO, C$^{18}$O, C$^{17}$O, $^{13}$C$^{18}$O) allow us to directly trace a wide range of gas masses in the cavity between a full gas disk ($\delta_{\rm gas}=1$) down to the detection limit of $^{12}$CO at $\delta_{\rm gas} \sim (0.3 - 3) \times 10^{-4}$. With the ALMA sensitivity we will be able to routinely trace gas masses in the cavity of transition disk in nearby star forming regions down to a fraction of the Earth mass ($\sim 0.2$ M$_{\rm Earth}$). This value is reached with a $5\sigma$ detection after a 1 hr on-source observation, thus lower masses can be traced with longer observations (Sect. \ref{sec:tracegasmass}).
\item The presence of a dusty inner disk (``pre-transition disk''), which shields the gap from  direct stellar irradiation, allows CO and H$_2$ to survive in the gap down to lower gas masses corresponding to $\delta_{\rm gas} \sim 10^{-6}$. In the absence of a dusty inner disk, the molecules can only survive for $\delta_{\rm gas} \gtrsim 10^{-4}$ (Sect. \ref{sec:innerdisk}). Differences in abundance and derived intensities between models with and without inner disk increase at lower values of $\delta_{\rm gas}$ (Sect \ref{sec:column}, \ref{sec:tracegasmass}).
\item High resolution observations with beam size corresponding to a fraction of the gap size are crucial to determine the amount of gas in the cavity (Sect. \ref{sec:needhighresol}). 
\item Observations of atomic fine structure lines of carbon ([\ion{C}{I}] and [\ion{C}{II}]) and oxygen ([\ion{O}{I}]) are less suited to measure the amount of gas in the cavity compared to low-$J$ lines of CO (Sect. \ref{sec:atomcarbon}, \ref{sec:column}, \ref{sec:carbonI}, and \ref{sec:herschel}).
\end{enumerate}

In extension to this work, which presents a parameter study, we will compare model results with ALMA observations (\citealt{vdMarel13}, Bruderer et al., in prep.) and also study the emission of other molecular tracers (e.g. of HCO$^+$ or CN).

\begin{acknowledgements}
I am grateful to Ewine van Dishoeck, Gregory Herczeg, Til Birnstiel, Cornelius Dullemond, Davide Fedele, Nienke van der Marel, Geoff Mathews, and the anonymous referee for stimulating discussions. I thank Ruud Visser for providing the routines to calculate PAH opacities. I acknowledge a stipend by the Max Planck Society.
\end{acknowledgements}

\bibliographystyle{aa}

\Online 

\clearpage

\begin{appendix}

\section{Improvements and changes of the model compared to \cite{Bruderer12}} \label{sec:app_mod}

\subsection{Continuum radiative transfer}\label{sec:app_cont}

The 3D Monte-Carlo radiative transfer method implemented in the model consists of two stages (see \citealt{Bruderer12} A.1. and A.2). In the first stage, the dust temperature is obtained using a method combining the algorithm by \cite{Bjorkman01} and \cite{Lucy99}. In order to improve the accuracy of the dust temperature in very optically thick midplanes and to speed up the calculation, we have replaced the modified random walk (MRW) approach (\citealt{Min09,Robitaille10}) by the solution of the diffusion equation. Following a suggestion by \cite{Pinte09}, photon packages are mirrored back from very optically thick regions in the midplane where opacities at the peak wavelength of the star are $\tau > 1000$. At this depth in the disk, the radiation field is close to isotropic and the mean intensity is similar to a black body of the local radiation field. Thus, the diffusion approximation may be employed. After all photon packages have been propagated, the dust temperature in the midplane is obtained from the solution of the diffusion equation 
\begin{equation} \label{eq:diffeq}
\nabla ( D \nabla T_{\rm dust}^4 ) = 0 \ , 
\end{equation}
where $D$ is the diffusion constant $D = 1 / (3 \rho \kappa_R(T_{\rm dust}))$, with the Rosseland mean mass extinction coefficient $\kappa_R(T_{\rm dust})$. The diffusion equation is also used for cells with low photon statistics (see \citealt{Min09}).

\begin{figure}[htb]
\includegraphics[width=1.0\hsize]{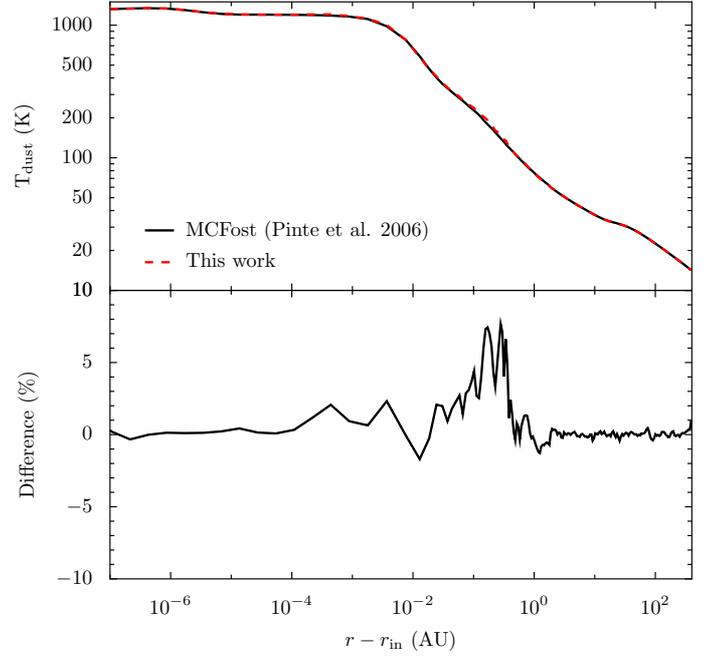}
\caption{Disk midplane dust temperature for the disk benchmark problem proposed in \citealt{Pinte09} ($\tau=10^5$). The result obtained from our code is given in the red dashed line, the results by MCFost (\citealt{Pinte06}) in black.}
\label{fig:bench_pinte}
\end{figure}

To verify the changes made in the code, we have re-run the benchmark tests suggested in \cite{Pinte09} (see Figure A.3 in \citealt{Bruderer12}). As an example, Figure \ref{fig:bench_pinte} shows a comparison of the midplane dust temperature obtained with our updated code and MCFost (\citealt{Pinte06}). The model with isotropic scattering and a midplane opacity of $\tau=10^5$ is shown. We find very good agreement with deviations comparable to those between other codes tested in the benchmark study. The main improvement to \cite{Bruderer12} is the reduced  calculation time combined with a better accuracy in regions of high opacity or low photon statistics, like regions shielded by a very opaque inner rim.

The second stage of the Monte-Carlo method calculates mean intensities (\citealt{Bruderer12}, A.2.). This part has only been used for FUV wavelengths before. We have updated the code to calculate mean intensities of the entire spectrum from UV to radio wavelengths. We typically use about 60 wavelength bins. The updated code also accounts for photons emitted by the dust. The fraction of photon packages emitted by the protostar, the dust and the interstellar radiation field or cosmic microwave background is chosen according to their relative contribution to the luminosity at one wavelength (e.g. \citealt{Pinte06,Harries11}).

The accuracy of the mean intensity calculated in the second stage can be verified by checking the consistency with the dust temperature obtained in the first stage. We compare the energy balance
\begin{equation}
\int \kappa_\nu J_\nu d\nu = 4 \pi \int \kappa_\nu B_\nu(T_{\rm dust}) d\nu
\end{equation}
using the dust temperature $T_{\rm dust}$ from the first stage and the mean intensity $J_\nu$ from the second stage and find agreement to within a few percent. The advantage of splitting the calculation in two stages is that the photons propagating in the first stage mainly sample the peak wavelength of the local radiation field to obtain an accurate dust temperature, while the photons in the second stage are distributed equally between the wavelengths to obtain a well sampled radiation field at all wavelengths. This is important for photodissociation, chemistry, and excitation.

\subsection{Molecular excitation} \label{sec:app_molexcit}

The main difference between \cite{Bruderer12} and the present model is the non-LTE calculation of the molecular/atomic excitation. In order to simplify the calculation, we chose to compute the molecular excitation in a 1+1D approach only accounting for the vertical and radial direction to obtain the escape probabilities. The rate coefficient for transitions between level $i$ and $j$ (Eq. A.17 in \citealt{Bruderer12}) is
\begin{equation}
P_{ij} = \left\{
\begin{array}{ll}
A_{ij}\epsilon_{ij}+B_{ij} \eta_{ij} \langle J_{ij} \rangle + C_{ij} & (E_i > E_j) \\
B_{ij} \eta_{ij} \langle J_{ij} \rangle + C_{ij} & (E_i < E_j) \ , \\
\end{array} \right.
\end{equation}
with the collisional rate coefficients $C_{ij}$, the Einstein A and B-coefficients $A_{ij}$ and $B_{ij}$, respectively and the level energy $E_i$. The continuum mean intensity $\langle J_{ij} \rangle$ is taken from the Monte-Carlo radiative transfer calculation. The escape probabilities in vertical direction  $\epsilon_{ij}$ and radial direction $\eta_{ij}$ are obtained from (e.g. \citealt{Avrett65}),
\begin{eqnarray}
\epsilon_{ij}(\tau_{ij}) &=& \frac{1}{2} K_2(\tau_{ij}) = \frac{1}{2} \int_{-\infty}^{\infty} \, \phi(x) \, {\rm E}_2(\tau_{ij}\phi(x))\, dx \\
\eta_{ij}(\tau_{ij}) &=& M_2(\tau_{ij}) = \int_{-\infty}^{\infty} \, \phi(x) \, \exp(-\tau_{ij} \phi(x)) \, dx \ .
\end{eqnarray}
with $\tau_{ij}$, the line center opacity in vertical or radial direction, respectively. The line profile is taken to be a Doppler line profile function $\phi(x) = \exp(-x^2)/\sqrt{\pi}$ and $E_2(x)$ is the second exponential integral. The expression for $\epsilon_{ij}$ corresponds to the escape probability of a slab, averaged over angle and line profile and the expression for $\eta_{ij}$ to the escape probability averaged over the line profile. In order to quickly evaluate the integrals in the above expressions, they are implemented using a Pad\' e-approximation which is accurate to better than 0.1 \%  for the whole opacity range used in the models.

The new approach, which is similar to other 1+1D disk models (e.g. \citealt{Gorti04}, \citealt{Woods09}, \citealt{Woitke09}) has the advantage of including the detailed continuum radiation field from the Monte-Carlo calculation for the radiative pumping of the molecules. Another advantage is that it avoids the iteration over the whole structure in order to solve the thermal-balance problem. This cuts down the calculation time of a single disk model by an order of magnitude down to of order 2 hours on a standard desktop computer. The accuracy of a similar approach has been tested by \cite{Kamp09} who find very small differences between this approach and a full radiative transfer calculation.

\subsection{Raytracing} \label{sec:app_raytrace}

The raytracing to obtain spectral cubes of the line images has been extended by the contribution of scattered continuum photons to the source function. Also, we have implemented a step-size adjustment along the rays, which avoids that the raytracer takes too small steps in regions where the line does not intersect with the frequency of the current channel due to the velocity profile. This method is similar to the technique implemented in RadLite (\citealt{Pontoppidan09}) and speeds up the raytracing considerably.

As an additional test for the raytracing to those reported in \cite{Bruderer12}, we have calculated the emission of a completely optically thin line from a molecule that is only abundant in the inner disk with strong velocity gradients. In this situation, the velocity integrated flux in a beam much larger than the disk can be calculated analytically by summing up the contribution of all cells. We have compared the results of the raytracer with the analytical results and find agreement to within a few percent, independent of the inclination of the disk. We note this is a meaningful benchmark test, as strong velocity gradients in the inner disk require an accurate step-size adjustment along the ray in order not to miss narrow lines (\citealt{Pontoppidan09}).

\subsection{Thermal balance} \label{sec:app_thermbalance}

Several small changes have been implemented into the thermal balance calculation: 

\subsubsection*{Line heating/cooling}
In order to also account for atomic heating through absorption of UV/optical photons followed by radiative and collisional decay, we have implemented larger model atoms into the non-LTE calculation. Table \ref{tab:atoms} provides a list of the implemented atoms and the spectroscopic and collisional rates used. Collisional coefficients with electrons have mostly been taken from the Chianti database (v7.0, \citealt{Dere97}). Collisional rates with atomic or molecular hydrogen are missing for \ion{Mg}{II}, \ion{Si}{II}, \ion{S}{II} and \ion{Fe}{II}. For \ion{Si}{II} and \ion{Fe}{II} we use estimated collisional rates by \cite{Sternberg95}, which however only connect the fine structure levels within the electronic ground state. 

The molecular coolants considered in the non-LTE calculation are $^{12}$CO, $^{13}$CO, C$^{18}$O, OH, H$_2$O, CN and HCN. For the isotope ratios, we take $^{12}$C/$^{13}$C=70, $^{16}$O/$^{18}$O=560 and $^{18}$O/$^{17}$O=3.2 (\citealt{Wilson94}). 

Molecular data has been taken from the LAMDA database (\citealt{Schoier05}) except for $^{12}$CO, where also ro-vibrational levels are accounted for. The  ro-vibrational collisional rates are obtained using the approach by \cite{Chandra01}, using the current rates from the LAMDA database for the vibrational ground state.

\begin{table}[tbh]
\caption{Considered atomic lines}
\label{tab:atoms}
\centering
\begin{tabular}{lllll}
Species & Levels & Lines & Spectroscopy & Coll. rates \\
\hline\hline
\ion{C}{I}   & 13 & 30  & S05, NIST     & S05 \\
\ion{C}{II}  & 8  & 11  & S05, NIST     & S05, Chianti\\
\ion{O}{I}   & 50 & 235 & S05, NIST     & S05, Chianti, St\"oH00\\
\ion{Mg}{II} & 13 & 32  & Chianti       & Chianti \\
\ion{Si}{II} & 24 & 79  & Chianti       & Chianti, SteDal95 \\
\ion{S}{II}  & 13 & 44  & Chianti       & Chianti \\
\ion{Fe}{II} & 84 & 497 & Chianti       & Chianti, SteDal95 \\
\hline
\end{tabular}
\tablefoot{S05$=$\cite{Schoier05},\\
St\"oH00$=$\cite{Stoerzer00},\\
SteDal95$=$\cite{Sternberg95},\\
NIST$=$\cite{Ralchenko09},\\
Chianti$=$\cite{Dere97} \& \cite{Landi12}}
\end{table}

\subsection*{Pumping of atomic oxygen by OH photodissociation}

We now also account for the population of the $^{1}$D levels of atomic oxygen by the photodissociation of OH,
\begin{equation}
{\rm OH} + \gamma_{\rm FUV} \rightarrow {\rm O}({}^{1}{\rm D}) + {\rm H} \ .
\end{equation}
Following \cite{vanDishoeck84}, we assume that $\sim$ 50 \% of OH photodissociations lead to O($^1$D). 

\subsection*{H$_2$ heating/cooling}

X-ray heating through H$_2$ ionization (\citealt{Glassgold73}, \citealt{Meijerink05a}) is now also accounted for. For the H$_2$ heating and cooling, we use a model with 15 vibrational levels including the effects of fluorescence and excited formation similar to \cite{Sternberg95}. Since the previously used rates by \citealt{Rollig06} are fits to this small model, the heating/cooling rates of the two approaches agree very well, except for the case of high temperatures where the 15 level model can cool more efficiently than the two level approach since higher vibrational levels are excited (see \citealt{Rollig06}).

\subsection{Chemistry} \label{sec:app_chem}

The chemical network is taken consistently with \cite{Bruderer12}, except that we use the H$_2$ formation rate by \cite{Cazaux04,Cazaux10}. An additional reaction for the hydrogen loss of a hydrogenated PAH by a photoreaction,
\begin{equation}
{\rm PAH:H} + \gamma_{\rm FUV} \rightarrow {\rm PAH} + {\rm H}
\end{equation}
is added. It is set to a rate of $10^{-8} G_0$ s$^{-1}$, with $G_0$ the FUV field in units of the interstellar radiation field (\citealt{LePage01}).

To solve for the steady-state abundances, we use an improved version of our globally convergent Newton-Raphson solver (\citealt{Bruderer12}). In case this method fails to converge, we use a time-dependent solver. Instead of the DVODE package (\citealt{Brown89}), we use the LIMEX package (\citealt{Ehrig99}) which was found to be more robust.

\subsection{Testing}

\begin{figure*}[!htb]
\sidecaption
\includegraphics[width=0.7\hsize]{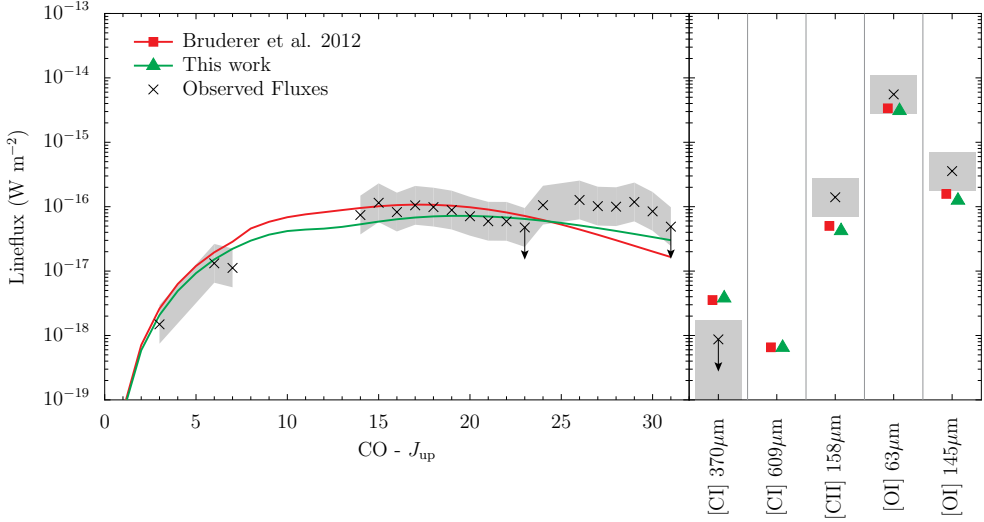}
\caption{Comparison of the model fluxes between the current model and the representative model of HD 100546 in \cite{Bruderer12}.}
\label{fig:comp_hd100546}
\end{figure*}

As a comparison, we have rerun the representative model of HD 100546 in \cite{Bruderer12} using the updated model. Line fluxes of both models together with observed line fluxes (\citealt{Sturm10}) are given in Figure \ref{fig:comp_hd100546}. While the atomic fine structure lines of C, C$^+$ and O agree to within 20 \% between the models, the mid-J and high-$J$ lines of CO show differences of up to a factor of two due to changes in the thermal balance. However, all conclusion made in \cite{Bruderer12} remain unchanged. As pointed out by \cite{Bruderer12}, the mid-$J$ and high-$J$ CO lines are particularly sensitive on the gas temperature.

\section{Intensity and flux conversions} \label{sec:app_fluxconv}

The conversion between intensity in units of Kelvin ($T_\nu$) and flux ($F_\nu$) requires the knowledge of the telescope beam. Assuming a Gaussian beam with full width at halve maximum of $\theta_{\rm FWHM}$, the solid angle of the beam is 
\begin{equation}
\Delta \Omega = \pi \frac{\theta_{\rm FWHM}^2}{4 \ln(2) }
\end{equation} 
Thus, the conversion between $T_\nu$ and $F_\nu$ is
\begin{equation}
F_\nu = \Delta \Omega \frac{ 2 \nu k}{c^2} T_\nu
\end{equation}
In units, this is
\begin{equation}
F_\nu \ \textrm{(Jy beam$^{-1}$)} = 8.18 \times 10^{-7} \cdot  (\nu \ \textrm{(GHz)} )^2 \cdot (\theta_{\rm FWHM} \cdot \textrm{('')})^2 \ T \textrm{(K)}
\end{equation}
For frequency/velocity integrated intensities it is 
\begin{equation}
\int F_\nu d\nu = \Delta \Omega \frac{2 \nu^3 k}{c^3} \int T_\nu d\textrm{v} \ ,
\end{equation}
where $d\nu/\nu = d\textrm{v}/c$ has been used.
Thus
\begin{eqnarray}
\int F_\nu d\nu\ \textrm{(W m}^{-2}\textrm{)} &=& 2.73 \times 10^{-29} \cdot  (\nu \ \textrm{(GHz)} )^3 \nonumber \\ &&  \cdot (\theta_{\rm FWHM} \cdot \textrm{('')})^2  \int T_\nu d\textrm{v} \ \textrm{(K km s}^{-1}\textrm{)} 
\end{eqnarray}

\section{Effect of different $R_{\rm gap}$ and $R_{\rm cav}$} \label{sec:app_diffgapcav}

\begin{figure}[htb]
\includegraphics[width=1.0\hsize]{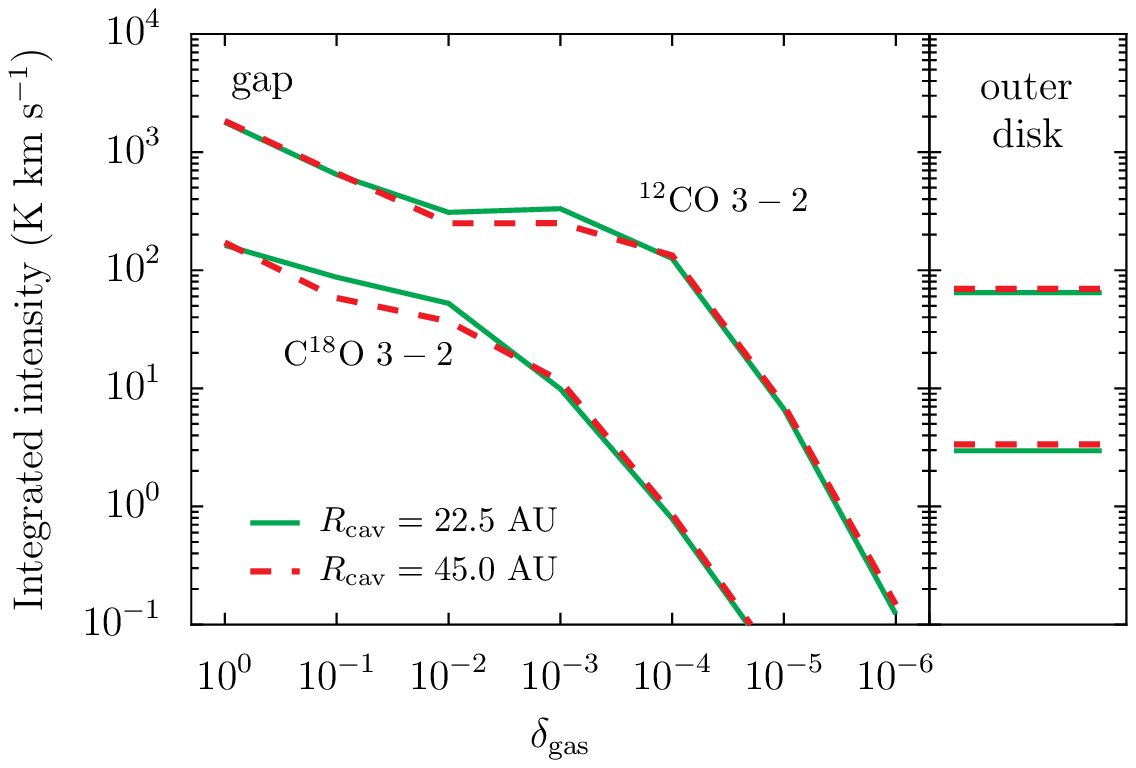}
\caption{The representative model with inner disk and dust gap from 10 AU to 45 AU compared to a model with a dust gap from 10 AU to 22.5 AU. Integrated intensities of $^{12}$CO $3-2$ and C$^{18}$O $3-2$ inside the gap ($r=16.25$ AU) and in the outer disk ($r=100$ AU) are shown.}
\label{fig:plot_flux_rcav}
\end{figure}

\begin{figure}[htb]
\includegraphics[width=1.0\hsize]{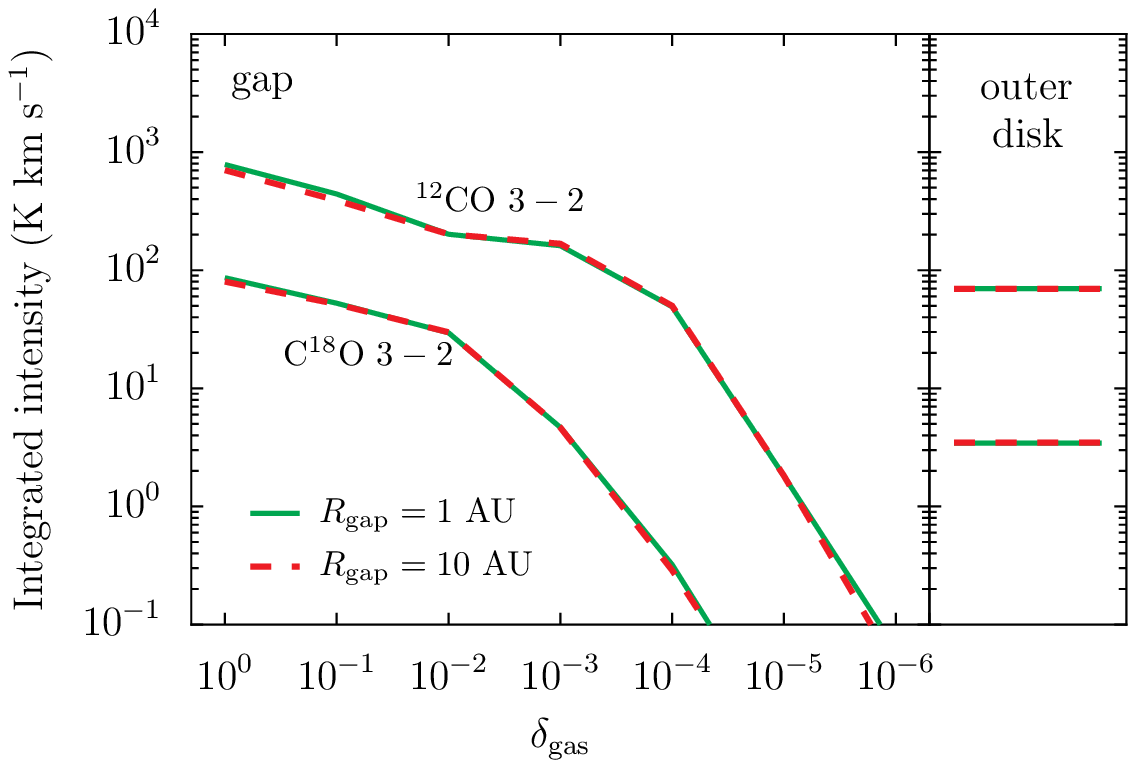}
\caption{The representative model with inner disk and dust gap from 10 AU to 45 AU compared to a model with a dust gap from 1 AU to 45 AU. Integrated intensities of $^{12}$CO $3-2$ and C$^{18}$O $3-2$ inside the gap ($r=28$ AU) and in the outer disk ($r=100$ AU) are shown.}
\label{fig:plot_flux_rgap}
\end{figure}

In order to study the effect of a different cavity size ($R_{\rm cav}$ in Figure \ref{fig:densstruct}a) and extent of the inner disk ($R_{\rm gap}$ in Figure \ref{fig:densstruct}a), we have calculated the representative model also with a dust gap extending from 1 AU to 45 AU ($R_{\rm gap}=1$ AU instead of 10 AU) and 10 AU to 22.5 AU ($R_{\rm cav}=22.5$ AU, instead of 45 AU). The resulting integrated intensities in the gap and outer disk are shown Figures \ref{fig:plot_flux_rcav} and \ref{fig:plot_flux_rgap} in the same way as Figure \ref{fig:plot_fluxdeltagas}. In Figure \ref{fig:plot_flux_rcav} we present the integrated intensity in the center of the smaller cavity, at 16.25 AU (see Sect. \ref{sec:tracegasmass}).

Models with a smaller cavity radius (smaller $R_{\rm cav}$) result in similar integrated intensities at a given radius inside the cavity, with differences $< 40$ \%. This results from the fact that a point at a certain radius is much more affected by positions at smaller radii than at larger radii, because of the stellar heating and photodissociating radiation coming from inside. We conclude that calculations with smaller cavities are necessary for a detailed modeling, but the trends found from the 45 AU gap also hold for smaller gaps.

Models with a smaller inner disk (smaller $R_{\rm gap}$) yield almost same integrated intensity profile inside the cavity, because the inner wall of the inner disk absorbs the stellar heating and photodissociating FUV radiation. This can be seen in Figure \ref{fig:repmod_phys}d, showing the local FUV radiation. We conclude that the size of the inner disk is not a key parameter, as long as its inner wall can shield the stellar radiation.

\section{Dependence on X-rays} \label{sec:app_xray}

\begin{figure*}[htb]
\includegraphics[width=0.95\hsize]{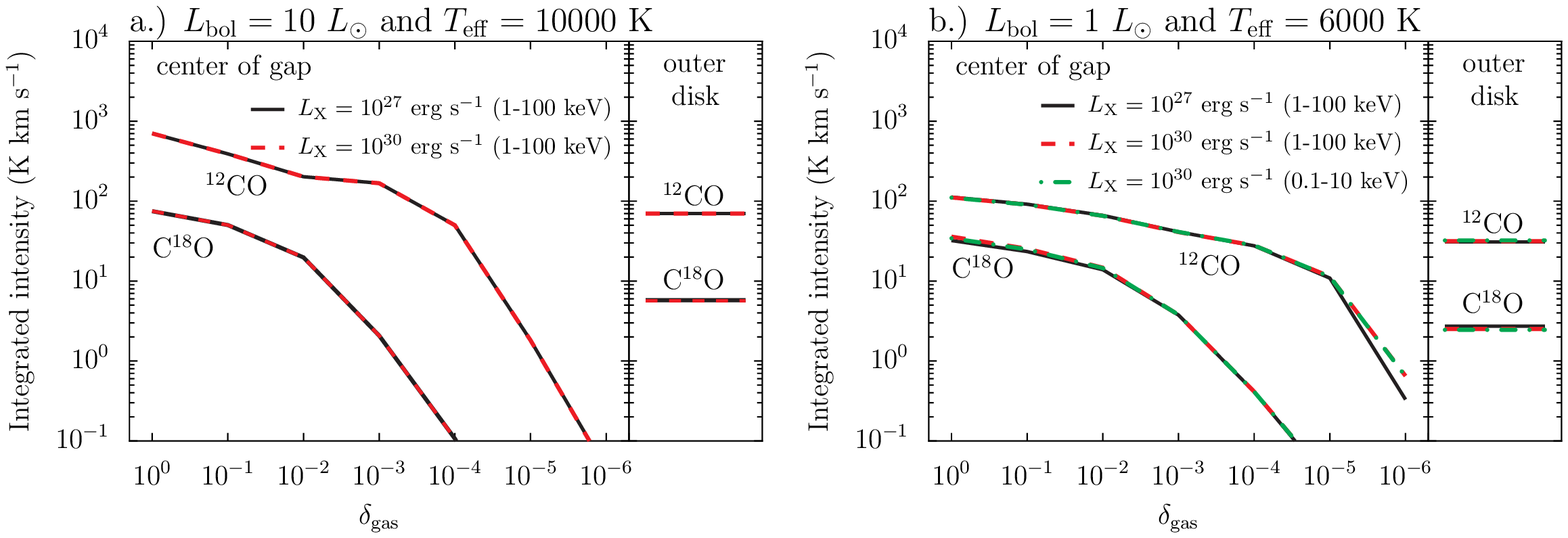}
\caption{Integrated intensities of $^{12}$CO $3-2$ at the center of the gap ($r=28$ AU) and in the outer disk ($r=100$ AU) for different X-ray luminosities ($L_{\rm X}$), stellar bolometric luminosities ($L_{\rm bol}$) and effective temperatures ($T_{\rm eff}$). The considered range of the X-ray spectrum is given in brackets (either 0.1-10 keV or 1-100 keV). Note that the lines in the left panel lie ontop of each other.}
\label{fig:plot_fluxdeltagas_xray}
\end{figure*}

The effect of X-ray radiation on the CO integrated intensity mainly depends on the X-ray luminosity ($L_{\rm X}$) compared to the FUV luminosity ($L_{\rm FUV}$). Figure \ref{fig:plot_fluxdeltagas_xray} shows the $^{12}$CO $3-2$ and C$^{18}$O $3-2$ integrated intensity for models with dusty inner disk present and $L_{\rm bol}=10$ $L_\odot$, $T_{\rm eff}=10000$ K or $L_{\rm bol}=1$ $L_\odot$, $T_{\rm eff}=6000$ K. The latter is the model with the highest $L_{\rm X}/L_{\rm FUV} \sim 1$ in our grid. However, even this model only shows a change of the $^{12}$CO integrated intensity in the gap with the X-ray luminosity, if the amount of gas in the cavity is low ($\delta_{\rm gas} \lesssim 10^{-5}$). This is because the heating of the upper layers in the gap is dominated by different effects (Sect. \ref{sec:gastemp}), while deeper in the disk, shielding by the inner disk is sufficient to render X-ray heating less important. The effect of X-rays does not change if the energy range of the thermal spectra is shifted to 0.1-10 keV instead of 1-100 keV. The rare isotopologue C$^{18}$O shows the same dependence on the X-rays as $^{12}$CO. Stellar X-rays are thus not an important parameter for the low-$J$ CO lines.

\end{appendix}

\end{document}